\documentclass[12pt, a4paper]{article}
\usepackage{graphicx}
\usepackage{amssymb}
\usepackage{amsmath}
\usepackage{bm}
\usepackage{color}
\usepackage{theorem}
\usepackage{subfigure}

\usepackage[sort&compress,numbers, merge]{natbib}

\definecolor{Orange}{cmyk}{0,0.61,0.87,0}
\definecolor{JungleGreen}{cmyk}{0.99,0,0.52,0}
\definecolor{OliveGreen}{cmyk}{0.64,0,0.95,0.40}
\definecolor{Brown}{cmyk}{0,0.81,1,0.60}
\definecolor{RoyalBlue}{cmyk}{0.71,0.53,0,0.12}

\newcommand{\Slash}[1]{{\ooalign{\hfil/\hfil\crcr$#1$}}}

\usepackage[colorlinks=true, linkcolor=OliveGreen, citecolor=RoyalBlue,
urlcolor=Brown]{hyperref}

\begin{document}

\begin{titlepage}

\begin{flushright}
FTPI-MINN-15/20 \\
UMN-TH-3433/15 \\
IPMU15-0063
\end{flushright}

\vskip 1.35cm
\begin{center}

{\LARGE
{\bf
Signatures of Leptoquarks at the LHC \\[5pt] and Right-handed Neutrinos
}
}

\vskip 1.2cm

Jason L. Evans$^a$
and
Natsumi Nagata$^{a,b}$

\vskip 0.4cm

{\it $^a$William I. Fine Theoretical Physics Institute, School of
 Physics and Astronomy, \\ University of Minnesota, Minneapolis, MN 55455,
 USA}\\ [3pt]

{\it $^b$Kavli IPMU (WPI), UTIAS, University of Tokyo, Kashiwa, Chiba
 277-8583, Japan}

\date{\today}

\vskip 1.5cm

\begin{abstract}

 In this paper, we argue that an extension of the Standard Model with a single
 leptoquark and three right-handed neutrinos can explain the excess in the
 first-generation leptoquark search at the LHC. We also find that when the
 leptoquark has similarly sized couplings to all three generations, it produces additional signals which will soon be tested in the second- and third-generation
 leptoquark searches, as well as in decay channels consisting of
 two mixed flavor leptons and two jets. If the leptoquark only couples
 to the first
 generation, on the other hand, two of the right-handed
 neutrinos need to be fairly degenerate in mass with the leptoquark while the other
 right-handed neutrinos mass should be much lighter. This hierarchical structure
 could explain dark matter and the baryon asymmetry of the
 Universe. These simple models may be regarded as benchmark models for
 explaining the excess, which can be tested in the next stage of the
 LHC running.

\end{abstract}

\end{center}
\end{titlepage}

\section{Introduction}

Leptoquarks are particles that have both baryon number and lepton number,
and appear in various beyond the standard model theories.
For example, some grand unified theories (GUTs), such as
SU(5) \cite{Georgi:1974sy}, $\text{SU}(4)_C\otimes\text{SU}(2)_L
\otimes \text{SU}(2)_R$ \cite{Pati:1974yy}, or SO(10)
\cite{Georgi:1974my, Fritzsch:1974nn} contain such particles. They are
also found in technicolor models \cite{Dimopoulos:1979es, Farhi:1980xs}
or other composite models \cite{Schrempp:1984nj}. Leptoquark
searches, therefore, provide a powerful test for these theories beyond
the Standard Model (SM).

Recently, the CMS Collaboration announced that they observed an excess
in the first-generation leptoquark search
\cite{CMS:2014qpa}. They analyzed the 8~TeV LHC data with an
integrated luminosity of 19.6~fb$^{-1}$, and looked for
first-generation leptoquarks in channels consisting of two
electrons and at least two jets ($eejj$), or an electron, a neutrino and
at least two jets ($e\nu jj$). An excess was found in both of
these channels. However, this excess was only found after optimizing the event selection
for leptoquarks with a mass of 650~GeV. The significance of the
excess is estimated to be $2.4\sigma$ and $2.6\sigma$ in the $eejj$ and
$e\nu jj$ channels respectively \cite{CMS:2014qpa}.
These excesses are, however, much smaller than those expected for a minimal
extension of the SM with a leptoquark. In fact, the analysis carried out
in Ref.~\cite{CMS:2014qpa} has excluded leptoquarks with masses
less than 1005 (845) GeV if the branching fraction of the leptoquarks
into an electron and a quark is equal to 1 (0.5). These constraints can be significantly relaxed, if the branching fraction to an electron and a jet is reduced.  Actually, this suggests that
leptoquarks have additional decay modes other than those with $eejj$ and $e\nu jj$ final states. Not surprisingly, leptoquarks may have a richer
structure than is found in these naive models.

Stimulated by this observation, various interpretations of the excess
have been proposed in the literature so far. For example, the authors in
Ref.~\cite{Bai:2014xba} introduce a coloron in addition to a leptoquark which could also explain the anomaly observed by the CMS Collaboration in the searches
for a right-handed $W$ gauge boson plus a heavy neutrino
\cite{Khachatryan:2014dka}. See also
Refs.~\cite{Heikinheimo:2014tba, Dobrescu:2014esa} for an explanation of both of these
excesses. Other leptoquark
models with a dark matter (DM) candidate are considered in
Refs.~\cite{Queiroz:2014pra, Allanach:2015ria}. These models may also explain
the excess in the two opposite-sign same-flavor
leptons with greater than two jets and missing transverse energy
\cite{CMS:2014jfa} channel. In supersymmetric theories, the existence of the
$R$-parity violating interactions may induce the observed excess;
this possibility is discussed in Refs.~\cite{Chun:2014jha,
Allanach:2014lca, Allanach:2014nna}.

In this paper, we study another, in fact quite simple, model which can explain the
excess in the first-generation leptoquark searches. This model consists of adding
a single leptoquark and three right-handed neutrinos to the SM. In contrast to previous work, we only consider
renormalizable interactions. As will be seen below, this model can actually
account for the observed excess. It turns out that when the leptoquark
couples to all three generations, signals in the second- and third-generation leptoquark searches are expected to appear in the near future.
Furthermore, decay modes containing a pair of mixed flavor
leptons and two jets are also expected.
If the leptoquark only couples to the first-generation fermions, on the
other hand, two of the right-handed neutrinos masses are required to be somewhat degenerate
with the leptoquark mass. The remaining right-handed
neutrino should be lighter than the electroweak scale.
This hierarchical structure in the right-handed neutrino sector
is quite interesting since the lightest right-handed neutrino can be a DM
candidate while the other two could help explain the baryon asymmetry of the
Universe (BAU). We will briefly discuss these possibilities.

\section{Model}
\label{sec:model}

\begin{table}[t]
\caption{Quantum numbers of scalar leptoquarks and possible
 renormalizable interactions containing leptoquarks. We follow the
 notation in Ref.~\cite{Agashe:2014kda}.}
\label{table:leptoquarks}
 \begin{center}
\begin{tabular}{c|cl}
\hline
\hline
 Leptoquarks & $\text{SU(3)}_C\otimes \text{SU(2)}_L\otimes
 \text{U(1)}_Y$ & Renormalizable couplings \\
\hline
$S_0$ & $({\bf 3}, {\bf 1}, -1/3)$ & $S_0 Q^\dagger_L L^\dagger_L$, $S_0
	 u_R^\dagger e_R^\dagger $, $S_0 Q_LQ_L$, $S_0 u_R d_R$\\
$\widetilde{S}_0$ & $({\bf 3}, {\bf 1}, -4/3)$ & $\widetilde{S}_0 d_R^\dagger
	 e_R^\dagger$, $\widetilde{S}_0 u_R u_R$ \\
$S_1$ & $({\bf 3}, {\bf 3}, -1/3)$ & $S_1 Q^\dagger_L L^\dagger_L$, $S_1
	 Q_LQ_L$ \\
$S_{1/2}^\dagger$ & $({\bf 3}, {\bf 2}, +7/6)$ & $S^\dagger_{1/2}
	 Q^\dagger_L e_R$, $S^\dagger_{1/2}u_R^\dagger L_L^{}$\\
$\widetilde{S}_{1/2}^\dagger$ & $({\bf 3}, {\bf 2}, +1/6)$ &
	 $\widetilde{S}^\dagger_{1/2} d^\dagger_R L_L^{}$ \\
\hline
\hline
\end{tabular}
\end{center}
\end{table}

To begin with, let us describe our model with a 650~GeV leptoquark,
which could account for the excess recently observed by the CMS Collaboration
\cite{CMS:2014qpa}. A comprehensive list of
leptoquarks\footnote{For a review, see the ``LEPTOQUARKS'' section
written by S.~Rolli and M.~Tanabashi in Ref.~\cite{Agashe:2014kda}. } is
found in Refs.~\cite{Buchmuller:1986zs, Davies:1990sc}. In this paper,
we focus on scalar leptoquarks since massive vector particles require
additional complexity in order to explain their masses, {\it i.e.}, the Higgs
mechanism. We list scalar leptoquark models in
Table~\ref{table:leptoquarks}. Among them, the first three types of leptoquarks
cannot be assigned a definite baryon ($B$) and lepton ($L$) number such
that the interactions of the leptoquarks conserve these charges
\cite{Nieves:1981tv}, since they have both leptoquark and
diquark couplings. In these cases, the tree-level
exchange of leptoquarks induces proton decay \cite{Nieves:1981tv,
Dorsner:2012nq, Arnold:2012sd, Arnold:2013cva}, which is stringently
constrained by experiments.\footnote{In the case of $\widetilde{S}_0$,
additional $W$-boson exchange is required for proton decay to occur,
since the $\widetilde{S}_0 u_R u_R$ interaction should include either a
charm or top quark due to the antisymmetry in the color indices. The
proton decay bound is still severe in this case. } For this reason,
we do not consider these possibilities in this paper. For the other two
leptoquarks, $S_{1/2}^\dagger$ and $\widetilde{S}_{1/2}^\dagger$, there
exist baryon-number violating dimension-five operators which are again
problematic for proton decay. As discussed in Ref.~\cite{Arnold:2013cva},
these operators can be forbidden by a $\mathbb{Z}_3$ symmetry under which each field transforms as $\psi \to
\exp[2\pi i(B-L)/3] \psi$. This symmetry could be a remnant of the
U(1)$_{B-L}$ symmetry \cite{Krauss:1988zc, Ibanez:1991hv,
*Ibanez:1991pr, Martin:1992mq}, which can be realized naturally in
GUTs.\footnote{For instance, the vacuum expectation
value (VEV) of a ${\bf 672}$ of SO(10) spontaneously breaks the
U(1)$_{B-L}$ symmetry into the $\mathbb{Z}_3$ symmetry
\cite{DeMontigny:1993gy,Mambrini:2015vna}.} Notice that this
$\mathbb{Z}_3$ symmetry cannot suppress the tree-level proton decay for
$S_0$, $\widetilde{S}_0$, and $S_1$, since the renormalizable
interactions given in Table~\ref{table:leptoquarks} conserve $B-L$. In
what follows, we will assume that the leptoquarks $S_{1/2}^\dagger$ and
$\widetilde{S}_{1/2}^\dagger$ do not lead to proton decay problems and focus on these two possibilities.

As is often the case for a new scalar particle coupling to
quarks and leptons, the addition of a leptoquark will, in general, cause
flavor/CP problems. New interactions of  a leptoquark with quarks
and leptons may induce flavor-changing neutral currents (FCNC) and/or
charged lepton flavor violation (CLFV), which are severely
restricted by low-energy precision experiments. Furthermore, leptoquarks
can also contribute to the electron and muon anomalous magnetic dipole
moments, and the electric dipole moments of quarks and leptons; again
the contribution should be small enough to be consistent with
current experiments. For the existing constraints on leptoquarks
from low-energy precision experiments, see
Refs.~\cite{Shanker:1982nd, Leurer:1993em, Davidson:1993qk, Gabrielli:2000te,
Mahanta:2001yc, Cheung:2001ip, Benbrik:2008si, Saha:2010vw,
Sahoo:2015wya, Varzielas:2015iva}. These bounds are especially severe for leptoquarks which
couple to both left- and right-handed quarks simultaneously, like
$S_{1/2}^\dagger$ in Table~\ref{table:leptoquarks}
\cite{Shanker:1982nd}. For this reason,
$\widetilde{S}_{1/2}^\dagger$ is more promising. Although it may be possible to construct a viable
model with $S_{1/2}^\dagger$, in this paper, we focus on
$\widetilde{S}_{1/2}^\dagger$, and for brevity denote it by $S$ in the following
discussion.

As seen in Ref.~\cite{CMS:2014qpa}, the CMS Collaboration observed slight
excesses in both the $eejj$ and the $e\nu jj$ channels. It may seem problematic that we consider the
$\widetilde{S}_{1/2}^\dagger$ leptoquark since it can only give rise to
the $eejj$ or $\nu\nu jj$ modes and not the $e\nu jj$ mode. Furthermore, the result in
Ref.~\cite{CMS:2014qpa} indicates that leptoquarks have
additional decay modes beyond those which produce the observed $eejj$ and
$e\nu jj$ final states. This motivates us to
consider the possibility that the leptoquarks couple to other particles that are singlet under $\text{SU}(3)_C\otimes
\text{U}(1)_{\text{EM}}$. With these additional couplings, it is possible to obtain the
$e\nu jj$-like events from decays to this singlet particle which yields a missing
transverse energy signal. In addition, these singlet particles allow us to hide
the signature of leptoquarks if their masses are somewhat degenerate with
the leptoquark mass, since the jets produced from the leptoquark decays become too
soft to be detected. These additional hidden decay modes
also result in a reduction of the branching fraction of the leptoquark to other states. Here, we will consider a singlet fermion which is coupled to the leptoquark through renormalizable interactions only.\footnote{In
Refs.~\cite{Queiroz:2014pra, Allanach:2015ria}, on the other hand,
leptoquarks are assumed to interact with a DM candidate through
non-renormalizable interactions.} The additional interactions of the leptoquarks are restricted by the SM gauge symmetries. The SM gauge symmetries indicate that
these interactions should contain a quark field since leptoquarks are
colored. The singlet fields should be fermionic in order to form a renormalizable
Lorentz scalar from a leptoquark, a quark, and a singlet field. As we
will see shortly, the leptoquark $S$ actually has such an
interaction with a singlet.\footnote{Notice that $S_{1/2}^\dagger$ does not couple to singlet
fields via renormalizable interactions. In this sense,
$\widetilde{S}_{1/2}^\dagger$ is the only candidate for our
considerations. However, it does still offer the possibility of hidden
decays to other generations. \label{shalfdagger}}

The leptoquark $S$ has $B=1/3$ and $L=-1$ in order for the model to preserve $B$ and $L$. Thus, unless the leptoquark coupling with the singlet violates lepton number, the singlet fermions should
have $|L|=1$. This charge assignment implies that these singlet fermions can be identified as right-handed neutrinos $\nu_{R}$. Indeed the
presence of right-handed neutrinos is well motivated by the observation
of neutrino oscillations, which indicates neutrinos are massive. Based on
the above discussion, we consider an
extension of the SM to which a leptoquark $S$ and three right-handed
neutrinos $\nu_{R_i}$ ($i=1,2,3$) are added.

The relevant interactions
for our discussion are then given by:
\begin{align}
 {\cal L}_{\text{int}}=& -\lambda_{ij} \epsilon_{\alpha\beta}
 (\overline{d_i})_a P_L(L_j)^\alpha S^{a\beta} -h_{ij}
 (\overline{Q_i})_{a\alpha} P_R \nu_j S^{a\alpha} \nonumber \\
 &-y_{ij} \epsilon_{\alpha\beta}\overline{\nu_i}P_L (L_j)^\alpha H^\beta
-\frac{1}{2}M_i \overline{\nu^c_i} P_R \nu_i ~~~~+\text{h.c} ~,
\end{align}
where $i,j$ are the generation indices; $a$ represents the color index;
$\alpha, \beta$ are SU(2)$_L$ indices; $c$ indicates the charge
conjugation; $\epsilon_{\alpha\beta}$ is the antisymmetric tensor with
$\epsilon_{12} = -\epsilon_{21} =+1$; $P_{L/R}\equiv
(1\mp\gamma_5)/2$. For later use, we write the SU(2)$_L$ components of
the leptoquark $S^{a\alpha}$ as
\begin{equation}
 S^a =
\begin{pmatrix}
 S^a_u \\ S^a_d
\end{pmatrix}
~.
\label{eq:leptdoub}
\end{equation}
The relative size of the neutrino Yukawa couplings $y_{ij}$ and Majorana
masses $M_i$ are chosen such that the left-handed neutrinos have small masses consistent with the oscillation experiments. All of the interactions except for the Majorana mass terms
conserve both baryon and lepton number, and thus do not induce
rapid proton decay. The off-diagonal components of
$\lambda_{ij}$ are strongly restricted by flavor experiments, and
thus we assume the new
couplings introduced above are diagonal, \textit{i.e.}, $\lambda_{ij} =
\lambda_i \delta_{ij}$ and $h_{ij} = h_i \delta_{ij}$. Of course, this assumption can be relaxed within the
experimental limits.

\section{Leptoquark signature at LHC}

At the LHC, leptoquarks are produced through gluon fusion and
quark-antiquark annihilation. Their production cross sections
are predominantly determined by the strong interactions, and the new
coupling constants $\lambda_{i}$ and $h_{i}$ give only a negligible
contribution to the leptoquark production since they are assumed to be very small. The reason for this will be clarified below. The NLO computation of the leptoquark pair production cross
section is given in Ref.~\cite{Kramer:2004df}.

After being produced, $S_u$ decays into the $d_R e_L^c$ or $u_L \nu_R^c$ final
state, while $S_d$ decays into the $d_R \nu_L^c$ or $d_L \nu_R^c$ channel. The
partial decay widths of these decay modes are as follows:
\begin{align}
\Gamma(S_u \to d_{Ri}^{}e_{Li}^c ) &= \Gamma(S_d \to d_{Ri}^{} \nu_{Li}^c )
=\frac{|\lambda_{i}|^2M_{\text{LQ}}}{16\pi}  ~, \nonumber  \\
 \Gamma(S_u \to u_{Li}^{}\nu_{Ri}^c ) &= \Gamma(S_d \to d_{Li}^{} \nu_{Ri}^c )
=\frac{|h_{i}|^2M_{\text{LQ}}}{16\pi} \biggl[1-\frac{M_{R_i}^2}
{M_{\text{LQ}}^2}\biggr]^2 ~,
\end{align}
where $M_{\text{LQ}}$ and $M_{R_i}$ are the masses of the leptoquark $S$
and right-handed neutrinos, respectively.
A pair production of $S_u$ can yield both the $eejj$ and the $e\nu jj$
signals, while $S_d$ gives $\nu\nu jj$ signals only. Since
the first-generation leptoquark searches by the CMS Collaboration only
look for the former two signals, the production of $S_d$ is irrelevant
for our discussion.

In these searches, the event selection for both $eejj$ and $\nu ejj$ signals is optimized for each leptoquark
mass by imposing additional cuts beyond those referred to as the preselection cuts. A list of these cuts can be found in Ref.~\cite{CMS:2014qpa}.
After all the cuts are imposed, the number of the leptoquark events is estimated to be $125.85(58)$ and $37.22(37)$ in the $eejj$ and $e\nu jj$ channels, respectively, for $M_{\text{LQ}} =
650$~GeV. Since the NLO production cross section for a 650 GeV
leptoquark is determined to be $13.2$~fb \cite{Kramer:2004df}, the signal
acceptances for the leptoquarks are $48.6$\% and $28.8$\% for $eejj$ and
$e\nu jj$ \cite{Bai:2014xba} channels, respectively, for the integrated luminosity
19.6~$\text{fb}^{-1}$. On the other hand, 36 events are observed with
$20.49 \pm 2.14 (\text{stat})\pm 2.45(\text{syst})$ events expected in
the $eejj$, and in the $e\nu jj$ channel 18 events are detected with
$7.54 \pm 1.20 (\text{stat})\pm 1.07(\text{syst})$ events expected. Therefore, to
explain these excesses the branching fractions of the
$S_u\to d e^c$ and $S_u \to u\nu^c$ are required to be
\begin{align}
 \text{BR}(S_u\to d e^c) &\simeq 0.35 ~,\nonumber \\
 \text{BR}(S_u\to u \nu^c) &\simeq 0.20 ~.
\label{eq:brs}
\end{align}
Clearly, the leptoquarks must have additional decay channels.

\begin{figure}[t]
\begin{center}
\subfigure[BR($S_u\to \mu s$) vs. BR($S_u\to \tau b$)]
 {\includegraphics[clip, width = 0.48 \textwidth]{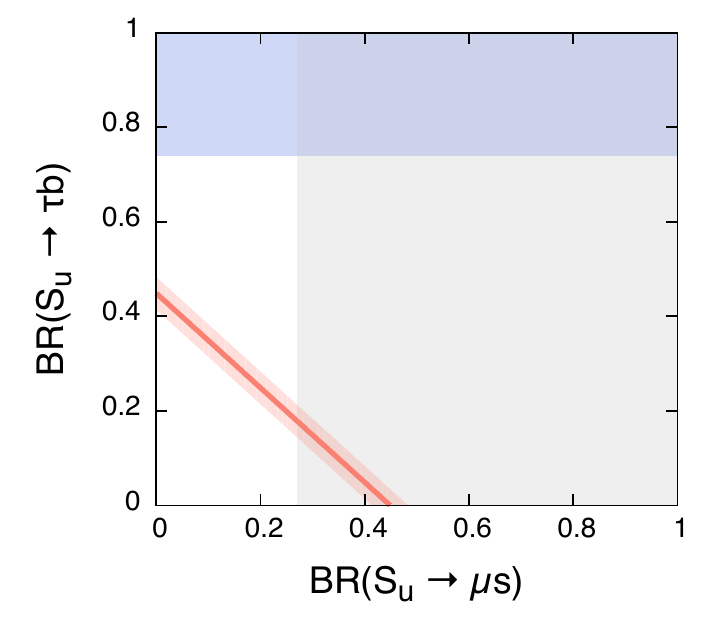}
 \label{fig:allgen}}
\hspace{-0.04\textwidth}
\subfigure[$ll^\prime jj$ channels]
 {\includegraphics[clip, width = 0.48 \textwidth]{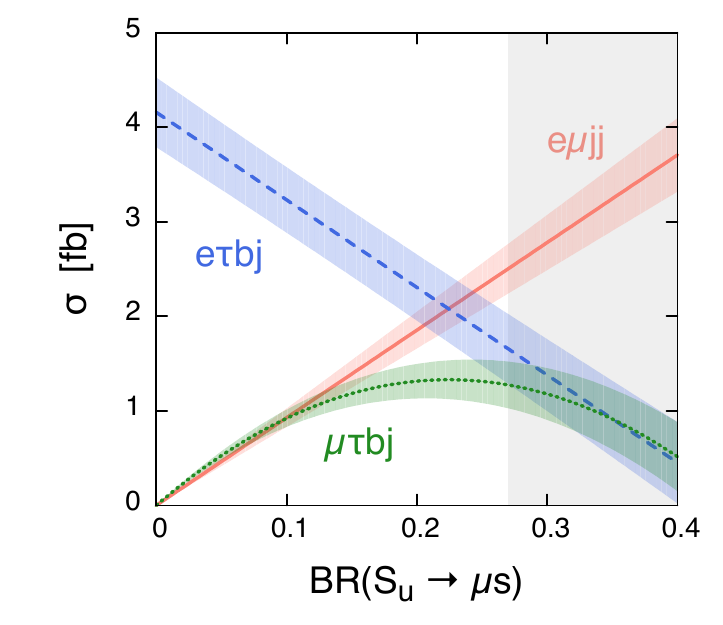}
 \label{fig:twoflv}}
\caption{(a) Branching fractions to the second and third generations
 that can explain the excess in Ref~\cite{CMS:2014qpa} are plotted in a red
 solid line with the corresponding band showing the uncertainty coming
 from the experimental errors in the event rates. Right gray-shaded region
 is excluded by the second-generation leptoquark search in the $\mu\mu
 jj$ channel \cite{CMS:zva}, while the upper blue shaded region is
 disfavored by the third-generation leptoquark search
 \cite{Khachatryan:2014ura}.
 (b) Scattering cross sections for the $ll^\prime jj$ modes as functions of
$\text{BR}(S_u \to \mu s)$. Red solid, blue dashed, and green dotted
 lines show the cross sections of the $e\mu jj$, $e\tau bj$, and $\mu
 \tau b j$ modes, respectively, with the corresponding bands indicating
 the uncertainties from the error in the observed event rates. The gray
 shaded region in the right side is excluded by the second-generation
 leptoquark search in the $\mu\mu jj$ channel \cite{CMS:zva}.
 The leptoquark mass is set to be $650$~GeV. }
\label{fig:3gen}
\end{center}
\end{figure}

One of the most straightforward and simple explanations of the additional decay
channels is that the leptoquark also decays into second- and
third-generations quarks and leptons. We refer to this scenario as Model
I in what follows.\footnote{As already mentioned in
footnote~\ref{shalfdagger}, $S^\dagger_{1/2}$ given in
Table~\ref{table:leptoquarks} may also explain the CMS excess in a
similar manner, though this type of leptoquark in general suffers from severe low-energy experimental
constraints. } For the
assumption $M_{R_i}\ll M_{\text{LQ}}$, the branching fractions of
$S_u$ are given by
\begin{align}
  \text{BR}(S_u\to d e^c) &= \frac{|\lambda_1|^2}{\sum_{i}|\lambda_i|^2
+\sum_{i}|h_i|^2}~,\nonumber \\[5pt]
 \text{BR}(S_u\to u \nu^c) &= \frac{\sum_{i}|h_i|^2}{\sum_{i}|\lambda_i|^2
+\sum_{i}|h_i|^2} ~.
\end{align}
Therefore, if we take the appropriate size for the couplings $\lambda_2$
and $\lambda_3$, we are able to explain the preferred branching
fractions found in Eq.~\eqref{eq:brs}. The decay modes of the
leptoquarks into the second- and third-generation quarks and leptons are
restricted by searches for leptoquarks which couple to only the second- or
third-generation quarks and leptons. For second-generation leptoquarks,
the CMS Collaboration studied the final states
consisting of two muons and at least two jets (the $\mu\mu jj$ channel)
and one muon,  and at least two jets, and missing transverse energy (the $\mu
\nu jj$ channel) with the 8~TeV 19.6~fb$^{-1}$ data \cite{CMS:zva}. The
former channel leads to the constraint $\text{BR}(S_u\to d\mu^c )\lesssim
0.27$ for a leptoquark of mass 650 GeV and the latter channel leads to
$\text{BR}(S_u\to d\mu^c )\lesssim 0.64$ assuming $\text{BR}(S_u \to
u\nu^c) = 0.20$ for the same leptoquark mass. They also searched for a
third-generation leptoquark with the 8~TeV 19.7~fb$^{-1}$ data set, and found the
bound on the branching fraction for a leptoquark decaying to a tau lepton and a bottom quark to be
$\text{BR}(S_u\to \tau b)\lesssim 0.74$ \cite{Khachatryan:2014ura}. The
decay mode of a leptoquark to a top quark and a neutrino can in
principle be constrained by the stop search, though the current constraint
is still very weak \cite{Chatrchyan:2013xna, CMS-PAS-EXO-13-010}. The
ATLAS Collaboration also searched for leptoquarks of these three
generations with their 7~TeV data set \cite{Aad:2011ch, ATLAS:2012aq,
ATLAS:2013oea}, though currently their constraints are weaker than the CMS
results.
In Fig.~\ref{fig:allgen}, we show the preferred branching fractions to the
second and third generations in a red solid line with the corresponding
band representing the uncertainty coming from the experimental errors in event
rates. Here, the right gray-shaded region is excluded by the second-generation leptoquark search in the $\mu\mu jj$ channel \cite{CMS:zva},
while the upper blue shaded region is disfavored by the third-generation
leptoquark search \cite{Khachatryan:2014ura}. We find that to explain
the excess in Ref.~\cite{CMS:2014qpa} the leptoquark must have a sizable branching fraction into the
third-generation fermions, $\text{BR}(S_u \to \tau b)\gtrsim 0.2$. This rather large branching fraction may be probed in the upgraded LHC experiments. Note that since
only the branching ratios are relevant to our discussion, the size of
the new couplings can be small. Thus, we need not
worry about non-perturbativity of the couplings nor the constraints
coming from single leptoquark production at HERA experiments \cite{Abramowicz:2012tg, South:2013fta}
and the LHC experiments \cite{CMS-PAS-EXO-12-043, Dorsner:2014axa, Mandal:2015vfa}.
Because the leptoquarks can decay to second- and third-generation quarks
and leptons, this model can be verified by searches for 
leptoquarks with final states consisting of different flavors of
charged leptons and two jets (the $ll^\prime jj$ channels with
$l\ne l^\prime$, and $l',l=e,\mu, \tau$). In Fig.~\ref{fig:twoflv}, we plot the
scattering cross sections for these final states as functions of
$\text{BR}(S_u \to \mu s)$. Here, the red solid, blue dashed, and green
dotted lines show the cross sections of the $e\mu jj$, $e\tau bj$, and
$\mu \tau b j$ modes, respectively, with the corresponding bands
indicating the uncertainties from the error in the observed event
rates in Ref.~\cite{CMS:2014qpa}. Again, the gray shaded region on the right side is excluded by the
second-generation leptoquark searches in the $\mu\mu jj$ channel
\cite{CMS:zva}. The leptoquark mass is set to be $650$~GeV. This figure shows
that the cross sections for the $ll^\prime jj$ channels can be several
fb, which are within the reach of the LHC experiments.

Here, we note that although we have assumed $M_{R_i}\ll M_{\text{LQ}}$
in this analysis, some or all of the right-handed neutrinos can have
masses of ${\cal O}(100)$~GeV without modifying the results too
much. This freedom is important when we discuss the
possibility of right-handed neutrino DM, as will be seen in
Sec.~\ref{sec:DMandBAU}.

Next, we consider the case where the leptoquark only couples to the
first-generation fermions, \textit{i.e.}, $\lambda_2 = \lambda_3 =0$. We
call this scenario Model II. For this
case, to realize the branching fractions in Eq.~\eqref{eq:brs}, we need
to hide some of the decays into the right-handed neutrinos. To accomplish this, we assume that some of the right-handed neutrinos are
degenerate in mass with the leptoquark. This makes the decays of the leptoquark to the degenerate right-handed neutrinos
invisible since jets in this decay process become too soft to be
detected. For example, the first-generation leptoquark searches in
Ref.~\cite{CMS:2014qpa} require that the leading jet should have a
transverse momentum larger than 125~GeV.  Thus if the mass differences
between the leptoquark and right-handed neutrinos are less than $\sim
100$~GeV, the jet from the decay of the leptoquark evades the
selection. From Eq.~\eqref{eq:brs}, we see that the invisible decays to right-handed neutrinos should have a
branching ratio of about $0.45$, which is twice as large as
the visible neutrino branching ratio $\text{BR}(S_u\to u\nu^c)$. This
observation is in line with two of the three right-handed neutrinos being somewhat degenerate in mass with the leptoquark. In this case, each branching ratio is
given by
\begin{align}
  \text{BR}(S_u\to d e^c) &= \frac{|\lambda_1|^2}{|\lambda_1|^2
+\sum_{i}|h_i|^2\bigl[1-\frac{M_{R_i}^2}
{M_{\text{LQ}}^2}\bigr]^2}~,\nonumber \\[5pt]
 \text{BR}(S_u\to u \nu^c) &= \frac{|h_1|^2}{|\lambda_1|^2
+\sum_{i}|h_i|^2\bigl[1-\frac{M_{R_i}^2}
{M_{\text{LQ}}^2}\bigr]^2} ~, \nonumber \\[5pt]
\text{BR(invisible)} &= \sum_{j=2,3}
\frac{|h_j|^2}{|\lambda_1|^2
+\sum_{i}|h_i|^2\bigl[1-\frac{M_{R_i}^2}
{M_{\text{LQ}}^2}\bigr]^2} \biggl[1-\frac{M_{R_j}^2}
{M_{\text{LQ}}^2}\biggr]^2 ~,
\end{align}
where we have assumed $M_{R_1} \ll M_{R_2}, M_{R_3}$. This gives $|h_1|\simeq
0.76 |\lambda_1|$. Moreover, if $|h_2|=|h_3|=h$ and $M_{\text{LQ}} - M_{R_2}
= M_{\text{LQ}} - M_{R_3} = 100$~GeV, then $h\simeq 2.8 |\lambda_1|$. At
present, the coupling $\lambda_1$ is a free parameter. However, this
prediction for the relation among the couplings can be important when
the lightest right-handed neutrino is the dominant component of DM in
the Universe, as we will
discuss in the next section. Also, $\lambda_1$ will be constrained in
this case.

Finally, we consider constraints on these models from the dijet plus
missing transverse energy $(jj+\Slash{E}_T)$ searches. After being produced,
100\% (4\%) of $S_d$ ($S_u$) pairs decay into two quarks and two
neutrinos giving the $jj+\Slash{E}_T$ events in the
detector. For a 650~GeV leptoquark decaying to $jj+\Slash{E}_T$, the current upper limit on the cross
section is about a few $\times 10$~fb
\cite{Khachatryan:2015vra, Aad:2014wea},\footnote{The ATLAS limit
given in Ref.~\cite{Aad:2014wea} only shows the results obtained
for a combined 2--6 jets plus missing energy search. This makes it rather difficult to ascertain the constraints on the production cross section for this mode alone. } which
is larger than the leptoquark production cross section. If some of the
right-handed neutrinos have similar masses to that of the leptoquark,
then the constraints are significantly relaxed. Anyways, we expect that
the LHC run II experiments can observe events in this mode. This is
a powerful consistency check of our models.

\section{Right-handed neutrino DM and the baryon asymmetry of the Universe}
\label{sec:DMandBAU}

In the previous section, it was found that the CMS measurement can be
explained by a leptoquark by adjusting the branching fractions
(couplings) of the leptoquark. However, the upper bound on the couplings
is independent of the CMS signal and a rather weak upper bound on the
couplings is placed by HERA \cite{Abramowicz:2012tg, South:2013fta}
and CMS \cite{CMS-PAS-EXO-12-043, Dorsner:2014axa, Mandal:2015vfa}
single-produced leptoquark searches. In fact, order-one couplings are still allowed.
The lower bound on the couplings of the leptoquark come from requiring
that the leptoquark decay be prompt. This amounts to having a decay
length shorter than about $1$~mm. If the decay length is longer than
$1$~mm, then the constraint from searches for displaced vertices can be
severe \cite{Aad:2015rba}. The total decay width for the
leptoquark discussed above is
\begin{equation}
\Gamma_S =\Gamma_{S_u}=
\Gamma_{S_d}\simeq
\sum\limits_i \left[|\lambda_i|^2+|h_i|^2\left(1-\frac{M_{R_i}^2}{M_{\rm
LQ}^2}\right)\right]\frac{M_{\rm LQ}}{16\pi}
=\frac{|\lambda_1|^2}{\text{BR}(S_u \to de^c)}\frac{M_{\rm LQ}}{16\pi}
 ~,
\end{equation}
for both Models I and II. Then, we find the decay length to be
\begin{eqnarray}
c\Gamma_{S}^{-1}\simeq 1~{\rm mm}\times \left(\frac{650~
 {\rm GeV}}{M_{\rm LQ}}\right) \times
 \left(\frac{7.3\times 10^{-8}}{\lambda_1}\right)^2
~.
\end{eqnarray}
As can be seen from
this expression, if the couplings are smaller than $10^{-7}$ their
decays can no longer be considered prompt. Finally, we find that the allowed range
for the leptoquark couplings is $10^{-7}\lesssim \lambda_1 \lesssim 1$,
and they are basically unrestricted by collider searches.\footnote{
Low-energy precision measurements, on the other hand, give fairly strong constraints on the leptoquark couplings. For
example, measurements of the parity-violating transition in cesium atoms
\cite{Wood:1997zq, Guena:2004sq} restrict the first-generation
leptoquark couplings so that $|\lambda_1| \leq 0.22$ \cite{Dorsner:2014axa}
for a 650~GeV leptoquark. The $Q_{\text{weak}}$ experiment
\cite{Androic:2013rhu}, which determines the weak charge of the proton by measuring the parity-violating asymmetry in
elastic electron-proton scattering, gives a similar constraint on this coupling. The most severe constraint for Model~I comes from $K_L \to \mu^- e^+$ decays. In fact, the constraint on the leptoquark couplings is $|\lambda_2 \lambda_1^*| < 0.9\times 10^{-5}$ \cite{Dorsner:2014axa,
Dorsner:2011ai}. As we will see, however, these are not the strongest constraints on the couplings of our model.}

The size of the leptoquark couplings $\lambda_i$ and $h_i$, however,
becomes important once we consider the phenomenology of right-handed
neutrinos. Right-handed neutrinos may play a significant role in both
particle physics and cosmology since they explain not only small neutrino
masses via the seesaw mechanism \cite{Minkowski:1977sc,*Yanagida:1979as,
*GellMann:1980vs, *Glashow:1979nm, *Mohapatra:1979ia}, but also DM and the
baryon asymmetry of the Universe. Indeed, it has been proposed that an
extension of the SM with three right-handed neutrinos can account for
all of these phenomena.  This scenario \cite{Asaka:2005an,*Asaka:2005pn} is called
the $\nu$ Minimal Standard Model ($\nu$MSM).\footnote{For a review, see
Ref.~\cite{Boyarsky:2009ix}. } It is therefore quite
interesting to see if adding a leptoquark  still gives a model which affords a
DM candidate and can generate the desired baryon asymmetry. We will see that the range for the leptoquark
couplings will be drastically reduced by the above requirements.

\subsection{Vanishing Leptoquark Couplings}

As a warm up, let us first consider the limit of vanishing $h_{i}$ and
$\lambda_i$. In this limit, the model considered in this article is
exactly that of the $\nu$MSM \cite{Asaka:2005an,*Asaka:2005pn}, for
which DM and baryogenesis can be explained. Later, we will discuss the
effects the leptoquark couplings have on the phenomenology of these models. As we will
see below, these couplings can drastically alter the situation.
In the $\nu$MSM, the lightest right-handed neutrino is assumed to have a mass of
${\cal O}(10)$~keV and to be the DM of the Universe. The masses of the other
two heavy right-handed neutrinos are quasi-degenerate and are of ${\cal
O}(0.1$--100)~GeV in order to generate the baryon asymmetry.
Both DM production and baryogenesis are non-equilibrium
processes. For this to be the case, it is assumed that no right-handed
neutrinos are generated during reheating. After reheating, the
right-handed neutrinos are generated by various scatterings of the
thermal bath.  Since the oscillations among different generations of right-handed neutrinos can be a non-equilibrium process which violates lepton number, a
lepton asymmetry is generated in the right-handed neutrinos among the different generations. The lepton sector includes extra CP
phases and thus offers new sources for CP-violation, which is required to produce a
baryon asymmetry. The asymmetry in the right-handed neutrinos is
converted to an asymmetry in the left-handed neutrinos via
oscillations. This asymmetry is then converted to a baryon asymmetry via
the sphaleron \cite{Kuzmin:1985mm}. Importantly, the right-handed
neutrinos do not reach equilibrium until after electroweak symmetry
breaking, when the sphaleron shuts off, preventing the generated asymmetry in the right-handed neutrinos, and thus in the left-handed neutrinos, from being washed out. Because the sphaleron process is inert when the right-handed
neutrinos reach equilibrium, the baryon asymmetry is preserved.

The production mode for the right-handed neutrinos is predominantly
through the scattering $t\bar t  \to h^* \to \nu_L N$ for the
$\nu$MSM, where $N$, $t$ and $h$ denote right-handed neutrinos, top
quark and the Higgs boson, respectively. When the temperature $T$ is
much higher than the masses of the particles participating in the
scattering process, this scattering leads to a production rate of
\begin{eqnarray}
\Gamma_N=n_{\rm EQ} \langle \sigma_{t\bar{t} \to \nu N }~ v\rangle
 \simeq 10^{-3}|y_{ij}|^2T  ~,
\end{eqnarray}
where $i\ne 1$, $n_{\text{EQ}}$ indicates the number density of a massless
particle and $\langle \sigma_{t\bar{t} \to \nu N} v\rangle$ is
the thermally averaged scattering cross section times relative
velocities. In the $\nu$MSM, the Yukawa couplings of the neutrinos with
the Higgs are quite small in order to explain small neutrino masses.  For the
heavier neutrinos they are of order $y_{ij}\lesssim 10^{-7}$ when $i\ne
1$ while $y_{ij} \lesssim 10^{-11}$ for $i=1$.\footnote{In the
presence of the leptoquark $S$, the Yukawa coupling can also be induced
at the loop level since the leptoquark couples to both left- and right-handed
neutrinos. This contribution is potentially dangerous since it could
generate Yukawa couplings which are too large. However, it turns out that this
contribution is smaller than that from the seesaw mechanism for the scenarios discussed
below, \textit{i.e.}, leptoquark Yukawa couplings ${\cal
O}(10^{-4})$.}  The heavier right-handed neutrinos will then come into
equilibrium much earlier than the lightest right-handed neutrino.  They
will come into equilibrium when
$\Gamma_N\sim H$ ($H$ is the Hubble expansion rate) which gives
\begin{eqnarray}
T_{\rm EQ}\simeq 10^{-3}|y_{ij}|^2\frac{M_P}{1.66 g_*^{1/2}}\sim
  \left(\frac{|y_{ij}|}{10^{-7}}\right)^2  \times
{\cal O}(10)
~{\rm GeV}~,
\end{eqnarray}
where $T_{\text{EQ}}$ is the temperature at which the heavier neutrinos
come into thermal equilibrium; $M_P =1.22\times 10^{19}$~GeV is the
Planck mass; $g_*$ is the
effective degrees of freedom for relativistic particles in the thermal bath.
A more detailed calculation shows that to ensure right-handed neutrinos are
out of thermal equilibrium before the sphaleron process shuts off, the
Yukawa couplings should be smaller than $2\times 10^{-7}$
\cite{Asaka:2005an,*Asaka:2005pn}, and thus the heavier right-handed
neutrino masses should be $M_{R_{2,3}}\lesssim 20$~GeV. Furthermore, this
scenario means that the largest asymmetry in the right-handed neutrinos,
and likewise in the left-handed neutrinos, is produced just before the
sphaleron shuts off. This larger asymmetry can then offset the small
mixing of the right- and left-handed neutrinos and produce the needed
baryon asymmetry.

DM in the $\nu$MSM is the lightest right-handed neutrino. It is well known
that a sterile neutrino with a mass of ${\cal O}(10)$~keV is a good warm
DM candidate \cite{Olive:1981ak, Peebles:1982ib}. The Yukawa couplings
of the lightest right-handed neutrino, if it is DM, need to be
sufficiently small so that its lifetime is longer than the age of the
Universe. The dominant decay mode of the right-handed neutrino is the
$N\to 3\nu$ channel. By evaluating the decay rate, one finds that an ${\cal
O}(10)$~keV right-handed neutrino naturally realizes a lifetime much
longer than the age of the Universe.
As it turns out, however, this is not the most severe restriction on
the couplings of the lightest right-handed neutrino if it is DM.
Its couplings are more severely constrained by searches for
diffuse X$/\gamma$ rays \cite{Pal:1981rm,Barger:1995ty,
Dolgov:2000ew,Abazajian:2001vt}. If the decay width is too large, it
will produce diffuse X$/\gamma$ rays which can be detected.  To avoid
these constraints, the lifetime needs to be longer than about $10^{28}$~s
which gives
\begin{equation}
\tau_{N_1\to\gamma\nu}=\Gamma^{-1}_{N_1\to\gamma\nu}
=\left(\frac{9\alpha
  G_F^2}{1024\pi^4}\sin^2(2\theta_1)M_{R_1}^5\right)^{-1}
\simeq  10^{28}\text{s}
\left(\frac{6.0\times 10^{-24}}{\sum_{i}|y_{i1}|^2}\right)\left(\frac{M_{R_1}}{\rm
 keV}\right)^3
 ~,
\end{equation}
with
\begin{equation}
\theta_1^2 =\sum\limits_i \frac{v^2|y_{i1}|^2}{M_{R_1}^2} ~,
\end{equation}
where $\alpha$ is the fine-structure constant; $G_F$ is the Fermi
constant; $v\simeq 246$~GeV is the Higgs VEV; $N_1$ denotes the lightest
right-handed neutrino DM.
As can be seen from these expressions, the decay is facilitated by a
mixing of the left- and right-handed neutrinos and a $W$-boson loop. By
inhibiting the oscillations between the right- and left-handed
neutrinos, this decay mode can be drastically suppressed. Suppressing
oscillation amounts to taking smaller values of $y_{ij}$.

Because the Yukawa couplings for the lightest right-handed neutrino are
extremely small, its production from top scatterings is suppressed and
therefore not enough DM is produced. This problem can be alleviated if a
large asymmetry is generated in the left-handed neutrinos which persists
to scales below electroweak symmetry breaking
\cite{Shi:1998km}.
This large asymmetry in the left-handed neutrinos can be produced via
neutrino oscillation of the two heavier right-handed neutrinos as they
freeze out or from resonantly enhanced decays \cite{Akhmedov:1998qx}.
For the oscillation production mechanism, the asymmetry is produced via
mixing among the right-handed neutrinos. When the mass scale associated
with mixing of the right-handed neutrinos, \textit{i.e.}, the differences of
the right-handed neutrino masses, are similar in size to the Hubble
parameter, an
asymmetry develops among the different species of the right-handed
neutrinos. This asymmetry is then converted to an asymmetry in the
left-handed neutrinos via mixing between the right- and left-handed
neutrinos. Since this oscillation occurs after freeze-out of the
right-handed neutrinos, the left-handed neutrinos do not oscillate back
to the heavier neutrinos. A similar mechanism is used to produce the
baryon asymmetry as the right-handed neutrinos are being generated by
scatterings of the thermal bath before equilibrium has been reached
\cite{Asaka:2005pn,Shaposhnikov:2008pf, *Canetti:2010aw, *Asaka:2010kk,
*Asaka:2011wq, *Canetti:2012vf, *Canetti:2012zc,
*Canetti:2012kh}. However, the production of the baryon asymmetry must
occur at a temperature above ${\cal O}(100)$ GeV before the sphaleron
shuts off, whereas the asymmetry needed for DM production is
produced below the scale where the sphaleron shuts off. If the asymmetry
in the left-handed neutrinos is produced via decays, the masses of the
right-handed neutrinos still need to be degenerate in order to enhance
the CP-violating parameter and produce a large enough asymmetry in the
left-handed neutrinos \cite{Pilaftsis:2003gt}. Once this asymmetry in
the left-handed neutrinos is produced, the thermal effects associated
with this non-zero chemical potential alter the neutrino mass matrix and
so alter the mixing relations between the left- and right-handed
neutrinos, like the Mikheyev-Smirnov-Wolfenstein effect \cite{Wolfenstein:1977ue}.  In fact, for some
temperature the mixing becomes of order one. If the temperature of the
Universe for which the mixing is order one is still greater than the
mass of the lightest right-handed neutrino, the left-handed neutrinos
can be converted into the lightest right-handed neutrino giving a
sufficient density to account for DM.

\subsection{Non-Vanishing Leptoquark Couplings}

For non-zero values of $h_{ij}$ and $y_{ij}$, things change drastically
since they couple the neutrinos with SM fields. These
additional couplings provide additional scattering processes which can
generate the right-handed neutrinos. The most important process among
them is the leptoquark-gluon scattering, $gS\to Q_L N$, which gives a
production rate of order
\begin{eqnarray}
\Gamma_N=n_{\rm EQ}\langle \sigma_{gS\to Q_L N} v\rangle
\sim 10^{-2}\alpha_s |h_{ij}|^2 T ~.
\end{eqnarray}
The equilibrium temperature for the right-handed neutrinos is then
\begin{eqnarray}
T_{\rm EQ}\sim 10^{-2}\alpha_s  |h_{ij}|^2\frac{M_P}{1.66 g_*^{1/2}}\sim
 10 \times\left(\frac{|h_{ij}|}{10^{-7}}\right)^2~\text{GeV}~.
\end{eqnarray}
If the couplings are taken as small as is allowed by the
prompt decay limit, we could get a situation similar to the $\nu$MSM for
baryogenesis.  Although this may work, it does significantly restrict
the parameter space of the model.

However, this production mechanism is very problematic for DM in Model II. Since the
LHC measurements require the coupling of each right-handed neutrino to
be of similar size, the amount of
the lightest right-handed neutrino produced will be similar to the other
right-handed neutrinos. This corresponds to a near thermal abundance for
the lightest right-handed neutrino, which will, in general, be too much DM.\footnote{
In Model I, it may be possible to suppress production of the lightest
right-handed neutrino by taking $h_1\ll h_{2,3}$.}  Since its decay
modes become severely inhibited for energies below the leptoquark mass,
the right-handed neutrinos produced will stick around and overclose the
Universe.

Below, we will discuss how the lightest right-handed neutrino could
still be DM in this scenario. However, first we discuss constraints on
the couplings of the lightest right-handed neutrino to leptoquarks if it
is indeed DM. These additional couplings will give additional
contributions to $\Gamma_{N_1\to \nu \gamma}$. Because the interaction
of the leptoquark with the neutrinos are vector like, this decay has a contribution coming from a loop involving a leptoquark and a quark \cite{Arcadi:2014dca}. A lifetime of the lightest
right-handed neutrino of order $10^{28}$~s, again required by the constraint
coming from diffuse X$/\gamma$ rays
\cite{Dolgov:2000ew,Abazajian:2001vt}, requires parameters of order
\cite{Arcadi:2014dca}\footnote{This constraint on the couplings is more
mild than that found in Ref.~\cite{Arcadi:2014dca}. This is due to the
lightest right-handed neutrino only coupling to the down-type quarks.}
\begin{eqnarray}
\Gamma_{N_1\to \nu_L \gamma}^{-1}\simeq
 \left(\frac{10^{-4}}{\lambda_1}\right)^2\left(\frac{10^{-4}}{h_1}\right)^2
 \left(\frac{M_{\rm LQ}}{650~{\rm GeV}}\right)^5\left(\frac{4 ~{\rm
  keV}}{M_{R_1}}\right)^3 \times  10^{28}~\text{s}~.
\label{eq:xraybound}
\end{eqnarray}
Fortunately, these couplings are much larger than those needed for the
prompt decay of the right-handed neutrinos. If the couplings lie right
at the upper edge allowed by this constraint, the leptoquarks with
right-handed neutrino DM could explain the 3.5~keV line observed in
X-ray spectra \cite{Bulbul:2014sua,*Boyarsky:2014jta}.

As can be seen from the previous paragraph, the constraints on the
couplings of right-handed neutrino DM to the leptoquark are not so
severe. The problems for producing the right-handed neutrino DM arise from thermal processes for $T\gtrsim M_{\rm
LQ}$. Below this temperature the leptoquarks decouple from the thermal
bath and have little effect on the cosmology. One way to circumvent this
problem is, therefore, to have a very low reheat temperature.  If the
reheat temperature $T_R$ is smaller than the leptoquark mass, the production
of the lightest right-handed neutrinos from $Sg\to N Q$
will be Boltzmann suppressed. This allows us to take larger values of
$\lambda$ and $h$ while keeping right-handed neutrinos from being over produced by the thermal bath.  For larger values of $h$ and $\lambda$, the scattering
process $\bar L d \to \bar Q N$ will become important.
Through this process, the right-handed neutrino DM can be non-thermally
produced. This is the same as the non-thermal equilibrium DM scenario discussed in
Refs.~\cite{Mambrini:2013iaa,Mambrini:2015vna}. In this scenario, the right-handed
neutrino DM is only weakly coupled to the thermal bath. This is due to the
small couplings $\lambda$ and $h$ and because the mediator of this process, the leptoquark, has a mass
somewhat larger than the reheating temperature. The lightest right-handed neutrinos are then produced by scatterings of the thermal bath.  Because of the small couplings and large leptoquark mass, the right-handed neutrinos never reach thermal equilibrium. However, for particular values of the couplings, leptoquark mass, and reheat temperature enough of the lightest right-handed
neutrinos can be produced to account for DM.

If the right-handed neutrinos come into equilibrium, they will overclose the Universe and could cause problems for galaxy formation
\cite{Olive:1981ak}. The right-handed neutrinos will come into
equilibrium when the production rate is equal to the Hubble
parameter. To guarantee that the lightest right-handed neutrino never
comes into equilibrium, we calculate the production cross
sections for $T<T_R < M_{\rm LQ}$ to leading order
\begin{equation}
 \langle \sigma_{{d}\bar{L}\to QN_1}~v\rangle
 =\sum\limits_i \langle \sigma_{{d}_i\bar{L}_i\to QN_1}~v\rangle\simeq \frac{3}{2\pi}N_CN_w\sum\limits_i
 |\lambda_ih_1|^2\frac{T^2}{M_{\rm
 LQ}^4} ~,
\end{equation}
%
where $N_C=3$ is the SU(3)$_C$ color factor and $N_w=2$ is
for the doublet factor of SU(2)$_L$.
The right-handed neutrino DM is not in thermal equilibrium if the
production rate is much smaller than the Hubble expansion rate. This
condition, $\langle \sigma_{QN_1\to d\bar{L}}v \rangle n_{N}^{\rm EQ} <
H$, is always satisfied if 
\begin{equation}
 \sum_{i}|\lambda_i h_i|^2 < 1.5 \times 10^{-14}
\left(\frac{g_*}{100}\right)^{\frac{1}{2}}
\left(\frac{M_{\text{LQ}}}{650~\text{GeV}}\right) ~,
\end{equation}
where we have used the inequality $T<T_R<M_{\text{LQ}}$. As can be seen, if $\lambda_i\simeq h_i \simeq
\lambda$ for $\lambda_i, h_i \neq 0$, then $\lambda < 3.5\times 10^{-4}$
satisfies this condition. For this case, the right-handed neutrino DM
cannot come into equilibrium after reheating.

After reheating, the number density for the right-handed neutrinos is effectively zero.  Since the number density is effectively zero, the annihilation rate of the right-handed neutrinos is negligible. The Boltzmann's equation for generating DM can be simplified to include only the production part \cite{Mambrini:2013iaa,Mambrini:2015vna}
\begin{eqnarray}
\frac{dY_N(x)}{dx}=
 \sqrt{\frac{\pi}{45}}\frac{g_s}{\sqrt{g_\rho}}M_PM_{N_1}
 \frac{\langle \sigma_{{d}\bar{L}\to QN_1}~v\rangle}{x^2} Y_{\rm EQ}^2 ~,
\end{eqnarray}
where $x=M_{R_1}/T$, $Y_N=n_N/s$, and $Y_{\text{EQ}}=n_{\text{EQ}}/s$;
$n_N$ is the number density of the lightest right-handed neutrino DM;
$s$ is the entropy of the Universe; $g_s$ and $g_\rho$ are the effective
degrees of freedom for the entropy and energy density respectively; $\langle \sigma_{{d}\bar{L}\to QN_1}~v\rangle$ is as above.
By integrating the above equation, we obtain the required reheat
temperature,
\begin{equation}
T_R\simeq 400\left(\frac{\Omega_{\text{DM}}
h^2}{0.12}\right)^{\frac{1}{3}}\left(\frac{g_*}{100}\right)^{\frac{1}{2}}
\left(\frac{M_{\rm LQ}}{650~{\rm GeV}}\right)^{\frac{4}{3}}\left(\frac{10~{\rm
keV}}{M_{R_1}}\right)^{\frac{1}{3}}\left(\frac{2}{N_G}\right)^{\frac{1}{3}}
\left(\frac{10^{-8}}{\lambda h}\right)^{\frac{2}{3}}~{\rm GeV} ~,
\end{equation}
where we set $g_s = g_\rho = g_*$. $N_G$ is the number of generations
participating in the production process, which is model dependent. $\Omega_{\text{DM}} h^2$ is the present DM density parameter.
This reheat temperature falls exactly where is needed for $N_1$ to be a good DM candidate. Namely,
it is just below the leptoquark mass so that the leptoquark interactions
are suppressed.  Furthermore, it is above the sphaleron process which means the
non-equilibrium dynamics could generate a baryon asymmetry via the
lepton asymmetry induced by oscillations in the heavier right-handed
neutrinos, as discussed above. This scenario is more plausible for Model~I, where we may take the heavier right-handed neutrino masses
to be ${\cal O}(1$--10)~GeV. The baryon asymmetry may also be generated
through decays after the right-handed neutrinos freeze out. Again, the asymmetry in the left-handed neutrinos is
converted to the baryon asymmetry through the sphaleron process.
Since the constraints from diffuse X/$\gamma$ rays only pertain to
the lightest right-handed neutrino, the leptoquark couplings could also
be adjusted to increase or decrease the baryon asymmetry as needed.

To determine the precise relic density of the right-handed neutrino DM and the
baryon asymmetry in our models, it is necessary
to numerically solve a set of coupled Boltzmann equations with thermal effects
adequately taken into account. This calculation is necessary to determine the leptoquark couplings with better accuracy. In
particular, a more detailed computation is needed to determine if these models can explain the 3.5~keV X-ray
line excess \cite{Bulbul:2014sua,*Boyarsky:2014jta}. This analysis will be left for future work.

\section{Gauge coupling unification}

It is widely known that the SM is not compatible with the minimal SU(5)
GUT \cite{Georgi:1974sy}. First, the
SM gauge couplings approach each other but never unify sufficiently
\cite{Georgi:1974yf}. Second, the
GUT scale is predicted to be too low to evade the current proton decay
constraints. Therefore, the simplest SU(5) GUT is ruled out.

Since the leptoquark $S$ is charged under the $\text{SU}(3)_C\otimes
\text{SU}(2)_L \otimes \text{U}(1)_Y$ gauge interactions, it can affect
the running of the SM gauge couplings. Thus, the presence of the
leptoquark may solve the problems mentioned above and revive the minimal
SU(5). Indeed, the leptoquark we consider can be embedded in a ${\bf
10}$, ${\bf 15}$, ${\bf 40}$, $\dots$ of SU(5). This allows us to
construct a GUT model where $S$ is a component of an irreducible
representation of SU(5).

At the two-loop level, the running of the SM gauge coupling constants $g_a$
($a=1,2,3$) is given by
\begin{equation}
 \mu \frac{d g_a}{d \mu}=\frac{1}{16\pi^2}b_a
^{(1)}g^3_a
+\frac{g_a^3}{(16\pi^2)^2}\biggl[
\sum_{b=1}^{3}b_{ab}^{(2)}g_b^2 -\sum_{k=u,d,e}c_{ak}~ {\rm
Tr}(f^\dagger_k f_k^{})
\biggr]~,
\end{equation}
where $g_1\equiv \sqrt{5/3}g^\prime $ and $f_k$ $(k =u,d,e)$ denotes the
Yukawa matrices of the SM fermions. The coefficients $b_a^{(1)}$,
$b_{ab}^{(2)}$, and $c_{ak}$ in the SM are given in
Ref.~\cite{Machacek:1983tz}. The change in the coefficient of the
beta function from $S$ is found to be
\begin{equation}
 \Delta b^{(1)} =
\begin{pmatrix}
 1/30 \\ 1/2 \\ 1/3
\end{pmatrix}
~, ~~~~~~
\Delta b^{(2)}=
\begin{pmatrix}
 1/150 & 3/10 & 8/15 \\
 1/10 & 13/2 & 8 \\
 1/15 & 3 & 22/3
\end{pmatrix}
~.
\end{equation}
Here, we neglect the effects of the leptoquark Yukawa couplings
since they are assumed to be very small.

\begin{figure}[t!]
\begin{center}
\subfigure[One leptoquark]
 {\includegraphics[clip, width = 0.48 \textwidth]{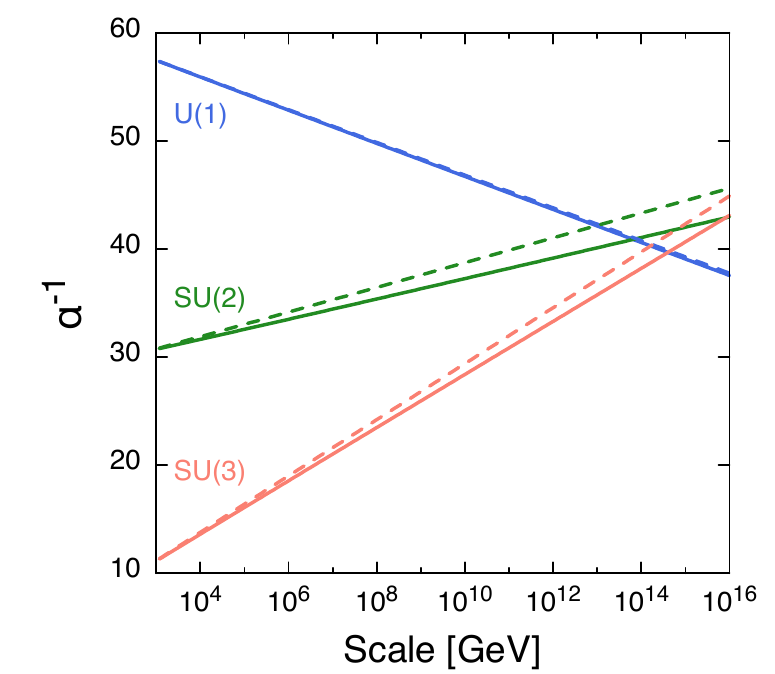}
 \label{fig:lq}}
\hspace{-0.04\textwidth}
\subfigure[Two leptoquarks]
 {\includegraphics[clip, width = 0.48 \textwidth]{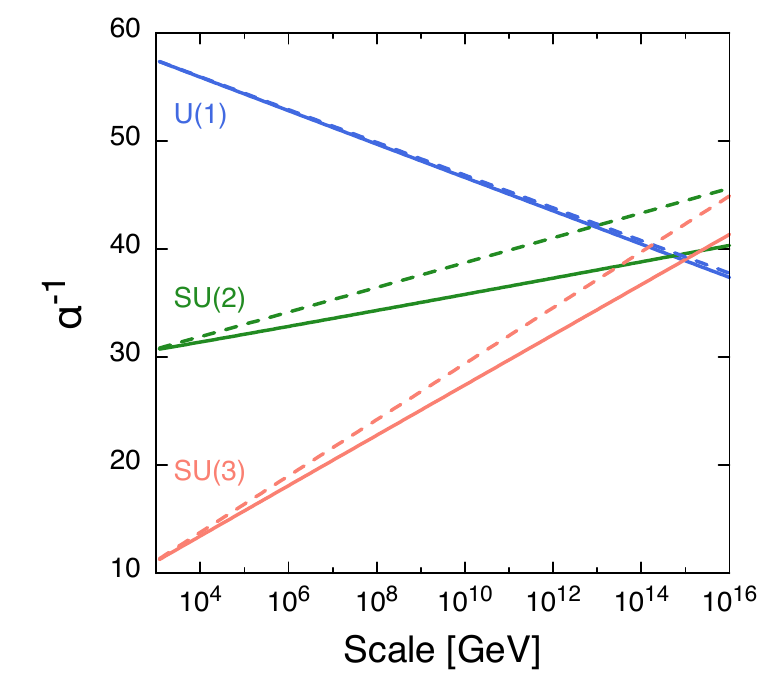}
 \label{fig:lq2}}\\
\subfigure[Two leptoquarks and one additional Higgs]
 {\includegraphics[clip, width = 0.48 \textwidth]{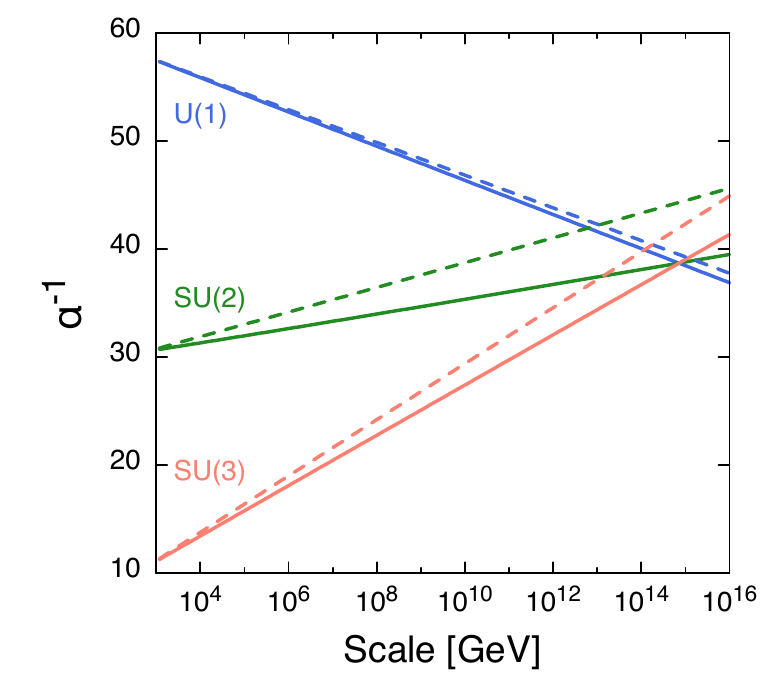}
 \label{fig:lq22hd}}
\caption{The running of the gauge couplings with (a) one leptoquark, (b) two
 leptoquarks or (c) two leptoquarks plus one additional Higgs doublet
 added to the SM are shown in solid lines. The SM running is shown in dashed lines for comparison. }
\label{fig:gcu}
\end{center}
\end{figure}

In Fig.~\ref{fig:lq}, we plot the running of the SM
gauge couplings with a leptoquark $S$ in solid lines. The blue, green,
and red lines show
$\alpha_a^{-1}$ ($a=1,2,3$), respectively, with $\alpha_a \equiv
g_a^2/(4\pi)$. The dashed lines show the running due to SM fields alone. In this
computation, we use the two-loop renormalization group equations given
above. However, we have not included threshold corrections at the GUT
scale since they are dependent on an unknown mass
spectrum. From this
figure, we see that although the presence of a leptoquark improves
gauge coupling unification, the deviation is still sizable and the unification scale is still rather low.

The situation is drastically improved, however, if there is an
additional leptoquark around the TeV scale. A leptoquark with a mass of
${\cal O}(1)$~TeV can easily evade the current LHC bounds. In
Fig.~\ref{fig:lq2}, we show the gauge coupling running for this
case. We also show the case where two leptoquarks and a Higgs
doublet\footnote{The contribution of a Higgs doublet to the
beta-function coefficients is given by
\begin{equation}
 \Delta b^{(1)} =
\begin{pmatrix}
 1/10 \\ 1/6 \\ 0
\end{pmatrix}
~, ~~~~~~
\Delta b^{(2)}=
\begin{pmatrix}
 9/50 & 3/10 & 0 \\
 9/10 & 13/6 & 0 \\
 0 & 0 & 0
\end{pmatrix}
~.
\end{equation}
} are
added to the SM in Fig.~\ref{fig:lq22hd}. This situation was already
noticed in Ref.~\cite{Murayama:1991ah}. As
can be clearly seen, gauge coupling unification in these two cases is far more precise. Accordingly, an extension of the SM with leptoquarks
provides an alternative and promising framework for an SU(5) GUT without
supersymmetry.

In the above cases, the unification scale is predicted to be $\sim
7\times 10^{14}$~GeV, which is still too low if one considers the bound from proton
decay. In the non-supersymmetric GUTs, proton decay is induced by
the exchange of the heavy SU(5) gauge bosons. The dominant
decay channel is then the $p\to e^+ \pi^0$ mode. The present experimental
limit on this channel is $\tau (p\to e^+ \pi^0)> 1.4\times
10^{34}$~years \cite{moriondsuperk}. To evade this constraint, we need to
assume that the heavy gauge boson mass $M_X$ is $\gtrsim 3\times
10^{15}$~GeV, which is much heavier than the unification scale found above. With
this disparity between scales, we expect sizable threshold corrections at the
GUT scale and thus the gauge couplings do not need to precisely unify \cite{Ellis:2015jwa}. In this sense, all of the models in
Fig.~\ref{fig:gcu} can be compatible with the SU(5) GUT framework and
thus they are quite promising. Anyways, all of these scenarios give a relatively
light SU(5) gauge boson, and thus future proton decay
experiments should detect proton decay in the $p\to e^+ \pi^0$
mode. This is a distinct prediction of these models compared with the SUSY SU(5) GUT
scenario. In the SUSY GUTs, the dominant decay channel is generically
the $p\to K^+ \bar{\nu}$ channel
\cite{Sakai:1981pk,*Weinberg:1981wj}. Additionally, since proton
decay in our scenario occurs via the gauge interactions, the various
decay modes are related by Cabibbo-Kobayashi-Maskawa (CKM) matrix elements. Thus, this model will give specific predictions for ratios such as $\text{BR}(p\to e^+ K^0)/\text{BR}(p\to e^+ \pi^0)$. As a
consequence of this, the leptoquark extension of the SM discussed in this paper
offers a simple way to achieve an SU(5) GUT model \cite{Murayama:1991ah,
Dorsner:2005fq} which can be tested in proton decay experiments in the near
future.

\section{BAU and DM From Additional Leptoquarks}
As was seen in the previous section, having two light leptoquarks can be
motivated by grand unification.  If we now include an additional
leptoquark in the model discussed above, we find additional freedoms
for generating the baryon asymmetry.  In fact, the gauge symmetries
allow the operator
\begin{eqnarray}
V_{B \!\!\!\!/} = q_{ijk}S_iS_jS_k H~,
\end{eqnarray}
where $i,j,k=1,2$. Although this operator breaks the discrete $B-L$
discussed above, it may still give an acceptably small proton decay because
the proton decay operators are induced at the two-loop level and the
couplings between the leptoquark and the SM particles are extremely
small. Another possible way to prevent proton decay would be to give the
second leptoquark a different charge under $B-L$ of $1/3$. With this
symmetry the above operator would still be allowed under the discreet
$B-L$, but the second leptoquark would be forbidden from coupling to the
SM particles alleviating proton decay.

This operator can be relevant for generating the baryon asymmetry. Since
the leptoquarks are scalars, it is possible that during inflation one of
the leptoquarks obtains a large VEV.  After inflation, the VEV of the
leptoquark will eventually return to zero.  However, since the couplings
$q_{ijk}$ can violate CP, the relaxation can generate angular momentum
in the leptoquark fields \cite{Affleck:1984fy}.  This angular momentum
violates baryon number and lepton number. The baryon number stored in
the leptoquark angular momentum generates a baryon asymmetry in the
quarks as the VEV decays. If conditions are right, this could be how the
baryon asymmetry was generated. The lepton number stored in the VEV
leads to the production of right-handed neutrinos when the VEV of the
leptoquark decays. This could be a non-thermal means of producing the
lightest right-handed neutrino. If the density is large enough, it could
be DM.


\section{Conclusion}

There are several unexplained signals lurking at the LHC.  One
particularly interesting signal is the excess in both $eejj$ and $e\nu
jj$ found in the CMS first-generation leptoquark searches
\cite{CMS:2014qpa}. The signals found in these searches rely on the
leptoquark having some decay paths which are hidden so the branching
fractions that produce the $eejj$ and $e\nu jj$ are sufficiently
small. If we assume the leptoquark has SM charges $({\bf 3},{\bf
2},1/6)$ in the basis $\text{SU(3)}_C\otimes \text{SU(2)}_L\otimes
\text{U(1)}_Y$, it will have the interactions $S\bar d L$ and $S\bar Q
\nu_R$ with generation indices suppressed. For this model, the simplest
way of explaining the hidden decays is to assume that the leptoquark
couples to all three generations. Since the current searches for
leptoquarks focus on leptoquarks that couple to each generation
individually, the hidden decay modes could merely be decays to other
generations. If this is the case, events with mixed-generation leptons
like $\tau ejj$ should be seen at the upgraded LHC.

On the other hand, if right-handed neutrinos are lighter than the
leptoquark, they could decay to right-handed neutrinos. If the two
heavier right-handed neutrinos are somewhat degenerate in mass with the
leptoquark, these decay modes would be excluded because the jets
produced in this event would be too soft.

For each of these scenarios, the lightest right-handed neutrino is a
viable DM candidate if its couplings to the leptoquarks are small enough
that its lifetime is longer than the age of the Universe.  If the
coupling of the right-handed neutrinos to the leptoquarks are small
enough, this model looks very similar to the $\nu$MSM
\cite{Boyarsky:2009ix} with similar production modes for the DM and
baryon asymmetry.  However, if these couplings are on the larger end of
the allowed range and the reheat temperature of the Universe is small
enough, the lightest right-handed neutrino DM can be a non-equilibrium
thermal DM candidate \cite{Mambrini:2013iaa,Mambrini:2015vna}.

With the leptoquark couplings small, the width of the leptoquark is
also small. Thus, we expect a narrow peak in the invariant mass distribution constructed from a
quark and a lepton final state. Since the statistics are low for this signal, we currently cannot
conclude whether the excesses observed in the CMS experiment are actually peaked or not. Future LHC experiments with more data
will be able to confirm or exclude this prediction.

The quality of the gauge coupling unification in the SM is made more
precise by the addition of a leptoquark.  It is even better with two
leptoquarks and an additional Higgs multiplet. Since proton decay in
these SM extensions is governed by interaction with the heavy gauge
bosons, the relative sizes of the different decay modes are determined by
the CKM matrix elements.  Since the mass of the heavy gauge bosons also
tends to be relatively light, these models make fairly precise predictions
for proton decay experiments.

The additional leptoquarks also give other possibilities for producing
the baryon asymmetry of the Universe. They allow for an Affleck-Dine
\cite{Affleck:1984fy} like production of the baryon asymmetry.

~\\~\\
{\it Note Added:} As we were about to post, we noticed 
Ref.~\cite{Dutta:2015dka} which also discusses an interpretation of
the CMS excess \cite{CMS:2014qpa} in terms of leptoquarks. However, this
model is quite different from our model.  They consider the leptoquark
$S_0$ instead of $\widetilde{S}_{1/2}^\dagger $ which we consider. They
also consider non-diagonal Yukawa couplings. The diagonality of our
couplings is advantageous because it does not have difficulty with rare
meson decays.

\section*{Acknowledgments}

The work of J.E. was supported in part by DOE grant DE-SC0011842 at the
University of Minnesota. The work of N.N. is supported by Research
Fellowships of the Japan Society for the Promotion of Science for Young
Scientists.


\bibliographystyle{JHEP}
\bibliography{ref}

\providecommand{\href}[2]{#2}\begingroup\raggedright\begin{thebibliography}{100}

\bibitem{Georgi:1974sy}
H.~Georgi and S.~Glashow, {\it {Unity of All Elementary Particle Forces}},
  {\em Phys.Rev.Lett.} {\bf 32} (1974) 438--441.

\bibitem{Pati:1974yy}
J.~C. Pati and A.~Salam, {\it {Lepton Number as the Fourth Color}},  {\em
  Phys.Rev.} {\bf D10} (1974) 275--289.

\bibitem{Georgi:1974my}
H.~Georgi, {\it {The State of the Art---Gauge Theories}},  {\em AIP Conf.Proc.}
  {\bf 23} (1975) 575--582.

\bibitem{Fritzsch:1974nn}
H.~Fritzsch and P.~Minkowski, {\it {Unified Interactions of Leptons and
  Hadrons}},  {\em Annals Phys.} {\bf 93} (1975) 193--266.

\bibitem{Dimopoulos:1979es}
S.~Dimopoulos and L.~Susskind, {\it {Mass Without Scalars}},  {\em Nucl.Phys.}
  {\bf B155} (1979) 237--252.

\bibitem{Farhi:1980xs}
E.~Farhi and L.~Susskind, {\it {Technicolor}},  {\em Phys.Rept.} {\bf 74}
  (1981) 277.

\bibitem{Schrempp:1984nj}
B.~Schrempp and F.~Schrempp, {\it {LIGHT LEPTOQUARKS}},  {\em Phys.Lett.} {\bf
  B153} (1985) 101.

\bibitem{CMS:2014qpa}
{\bf CMS} Collaboration, {\it {Search for Pair-production of First Generation
  Scalar Leptoquarks in pp Collisions at $\sqrt{s} = 8$ TeV}},  Tech. Rep.
  CMS-PAS-EXO-12-041, 2014.

\bibitem{Bai:2014xba}
Y.~Bai and J.~Berger, {\it {Coloron-assisted Leptoquarks at the LHC}},  {\em
  Phys.Lett.} {\bf B746} (2015) 32--36,
  [\href{http://arxiv.org/abs/1407.4466}{{\tt arXiv:1407.4466}}].

\bibitem{Khachatryan:2014dka}
{\bf CMS} Collaboration, V.~Khachatryan et~al., {\it {Search for heavy
  neutrinos and $\mathrm {W}$ bosons with right-handed couplings in
  proton-proton collisions at $\sqrt{s} = 8\,\text {TeV} $}},  {\em
  Eur.Phys.J.} {\bf C74} (2014) 3149,
  [\href{http://arxiv.org/abs/1407.3683}{{\tt arXiv:1407.3683}}].

\bibitem{Heikinheimo:2014tba}
M.~Heikinheimo, M.~Raidal, and C.~Spethmann, {\it {Testing Right-Handed
  Currents at the LHC}},  {\em Eur.Phys.J.} {\bf C74} (2014), no.~10 3107,
  [\href{http://arxiv.org/abs/1407.6908}{{\tt arXiv:1407.6908}}].

\bibitem{Dobrescu:2014esa}
B.~A. Dobrescu and A.~Martin, {\it {Interpretations of anomalous LHC events
  with electrons and jets}},  {\em Phys.Rev.} {\bf D91} (2015) 035019,
  [\href{http://arxiv.org/abs/1408.1082}{{\tt arXiv:1408.1082}}].

\bibitem{Queiroz:2014pra}
F.~S. Queiroz, K.~Sinha, and A.~Strumia, {\it {Leptoquarks, Dark Matter, and
  Anomalous LHC Events}},  {\em Phys.Rev.} {\bf D91} (2015) 035006,
  [\href{http://arxiv.org/abs/1409.6301}{{\tt arXiv:1409.6301}}].

\bibitem{Allanach:2015ria}
B.~Allanach, A.~Alves, F.~S. Queiroz, K.~Sinha, and A.~Strumia, {\it
  {Interpreting the CMS $\ell^+\ell^- jj E\!\!\!\!/_{\rm T}$ Excess with a
  Leptoquark Model}},  \href{http://arxiv.org/abs/1501.03494}{{\tt
  arXiv:1501.03494}}.

\bibitem{CMS:2014jfa}
{\bf CMS} Collaboration, C.~Collaboration, {\it {Search for physics beyond the
  standard model in events with two opposite-sign same-flavor leptons, jets,
  and missing transverse energy in pp collisions at $\sqrt{s} = 8$ TeV}},
  Tech. Rep. CMS-PAS-SUS-12-019, 2014.

\bibitem{Chun:2014jha}
E.~J. Chun, S.~Jung, H.~M. Lee, and S.~C. Park, {\it {Stop and Sbottom LSP with
  R-parity Violation}},  {\em Phys.Rev.} {\bf D90} (2014) 115023,
  [\href{http://arxiv.org/abs/1408.4508}{{\tt arXiv:1408.4508}}].

\bibitem{Allanach:2014lca}
B.~Allanach, S.~Biswas, S.~Mondal, and M.~Mitra, {\it {Explaining a CMS $eejj$
  Excess With $\mathcal{R}-$parity Violating Supersymmetry and Implications for
  Neutrinoless Double Beta Decay}},  {\em Phys.Rev.} {\bf D91} (2015) 011702,
  [\href{http://arxiv.org/abs/1408.5439}{{\tt arXiv:1408.5439}}].

\bibitem{Allanach:2014nna}
B.~Allanach, S.~Biswas, S.~Mondal, and M.~Mitra, {\it {Resonant slepton
  production yields CMS $eejj$ and $ep_Tjj$ excesses}},  {\em Phys.Rev.} {\bf
  D91} (2015) 015011, [\href{http://arxiv.org/abs/1410.5947}{{\tt
  arXiv:1410.5947}}].

\bibitem{Agashe:2014kda}
{\bf Particle Data Group} Collaboration, K.~Olive et~al., {\it {Review of
  Particle Physics}},  {\em Chin.Phys.} {\bf C38} (2014) 090001.

\bibitem{Buchmuller:1986zs}
W.~Buchmuller, R.~Ruckl, and D.~Wyler, {\it {Leptoquarks in Lepton - Quark
  Collisions}},  {\em Phys.Lett.} {\bf B191} (1987) 442--448.

\bibitem{Davies:1990sc}
A.~Davies and X.-G. He, {\it {Tree Level Scalar Fermion Interactions Consistent
  With the Symmetries of the Standard Model}},  {\em Phys.Rev.} {\bf D43}
  (1991) 225--235.

\bibitem{Nieves:1981tv}
J.~F. Nieves, {\it {Baryon and Lepton Number Nonconserving Processes and
  Intermediate Mass Scales}},  {\em Nucl.Phys.} {\bf B189} (1981) 182.

\bibitem{Dorsner:2012nq}
I.~Dorsner, S.~Fajfer, and N.~Kosnik, {\it {Heavy and light scalar leptoquarks
  in proton decay}},  {\em Phys.Rev.} {\bf D86} (2012) 015013,
  [\href{http://arxiv.org/abs/1204.0674}{{\tt arXiv:1204.0674}}].

\bibitem{Arnold:2012sd}
J.~M. Arnold, B.~Fornal, and M.~B. Wise, {\it {Simplified models with baryon
  number violation but no proton decay}},  {\em Phys.Rev.} {\bf D87} (2013)
  075004, [\href{http://arxiv.org/abs/1212.4556}{{\tt arXiv:1212.4556}}].

\bibitem{Arnold:2013cva}
J.~M. Arnold, B.~Fornal, and M.~B. Wise, {\it {Phenomenology of scalar
  leptoquarks}},  {\em Phys.Rev.} {\bf D88} (2013) 035009,
  [\href{http://arxiv.org/abs/1304.6119}{{\tt arXiv:1304.6119}}].

\bibitem{Krauss:1988zc}
L.~M. Krauss and F.~Wilczek, {\it {Discrete Gauge Symmetry in Continuum
  Theories}},  {\em Phys.Rev.Lett.} {\bf 62} (1989) 1221.

\bibitem{Ibanez:1991hv}
L.~E. Ibanez and G.~G. Ross, {\it {Discrete gauge symmetry anomalies}},  {\em
  Phys.Lett.} {\bf B260} (1991) 291--295.

\bibitem{Ibanez:1991pr}
L.~E. Ibanez and G.~G. Ross, {\it {Discrete gauge symmetries and the origin of
  baryon and lepton number conservation in supersymmetric versions of the
  standard model}},  {\em Nucl.Phys.} {\bf B368} (1992) 3--37.

\bibitem{Martin:1992mq}
S.~P. Martin, {\it {Some simple criteria for gauged R-parity}},  {\em
  Phys.Rev.} {\bf D46} (1992) 2769--2772,
  [\href{http://arxiv.org/abs/hep-ph/9207218}{{\tt hep-ph/9207218}}].

\bibitem{DeMontigny:1993gy}
M.~De~Montigny and M.~Masip, {\it {Discrete gauge symmetries in supersymmetric
  grand unified models}},  {\em Phys.Rev.} {\bf D49} (1994) 3734--3740,
  [\href{http://arxiv.org/abs/hep-ph/9309312}{{\tt hep-ph/9309312}}].

\bibitem{Mambrini:2015vna}
Y.~Mambrini, N.~Nagata, K.~A. Olive, J.~Quevillon, and J.~Zheng, {\it {Dark
  Matter and Gauge Coupling Unification in Non-supersymmetric SO(10) Grand
  Unified Models}},  {\em Phys.Rev.} {\bf D91} (2015) 095010,
  [\href{http://arxiv.org/abs/1502.06929}{{\tt arXiv:1502.06929}}].

\bibitem{Shanker:1982nd}
O.~U. Shanker, {\it {$\pi \ell 2$, $K \ell 3$ and $K^0 - \bar{K}^0$ Constraints
  on Leptoquarks and Supersymmetric Particles}},  {\em Nucl.Phys.} {\bf B204}
  (1982) 375.

\bibitem{Leurer:1993em}
M.~Leurer, {\it {A Comprehensive study of leptoquark bounds}},  {\em Phys.Rev.}
  {\bf D49} (1994) 333--342, [\href{http://arxiv.org/abs/hep-ph/9309266}{{\tt
  hep-ph/9309266}}].

\bibitem{Davidson:1993qk}
S.~Davidson, D.~C. Bailey, and B.~A. Campbell, {\it {Model independent
  constraints on leptoquarks from rare processes}},  {\em Z.Phys.} {\bf C61}
  (1994) 613--644, [\href{http://arxiv.org/abs/hep-ph/9309310}{{\tt
  hep-ph/9309310}}].

\bibitem{Gabrielli:2000te}
E.~Gabrielli, {\it {Model independent constraints on leptoquarks from rare muon
  and tau lepton processes}},  {\em Phys.Rev.} {\bf D62} (2000) 055009,
  [\href{http://arxiv.org/abs/hep-ph/9911539}{{\tt hep-ph/9911539}}].

\bibitem{Mahanta:2001yc}
U.~Mahanta, {\it {Implications of BNL measurement of $\delta a_\mu$ on a class
  of scalar leptoquark interactions}},  {\em Eur.Phys.J.} {\bf C21} (2001)
  171--173, [\href{http://arxiv.org/abs/hep-ph/0102176}{{\tt hep-ph/0102176}}].

\bibitem{Cheung:2001ip}
K.-m. Cheung, {\it {Muon anomalous magnetic moment and leptoquark solutions}},
  {\em Phys.Rev.} {\bf D64} (2001) 033001,
  [\href{http://arxiv.org/abs/hep-ph/0102238}{{\tt hep-ph/0102238}}].

\bibitem{Benbrik:2008si}
R.~Benbrik and C.-K. Chua, {\it {Lepton Flavor Violating $l\to l^\prime \gamma$
  and $Z\to l \bar{l}^\prime$ Decays Induced by Scalar Leptoquarks}},  {\em
  Phys.Rev.} {\bf D78} (2008) 075025,
  [\href{http://arxiv.org/abs/0807.4240}{{\tt arXiv:0807.4240}}].

\bibitem{Saha:2010vw}
J.~P. Saha, B.~Misra, and A.~Kundu, {\it {Constraining Scalar Leptoquarks from
  the K and B Sectors}},  {\em Phys.Rev.} {\bf D81} (2010) 095011,
  [\href{http://arxiv.org/abs/1003.1384}{{\tt arXiv:1003.1384}}].

\bibitem{Sahoo:2015wya}
S.~Sahoo and R.~Mohanta, {\it {Scalar leptoquarks and the rare B meson
  decays}},  {\em Phys.Rev.} {\bf D91} (2015) 094019,
  [\href{http://arxiv.org/abs/1501.05193}{{\tt arXiv:1501.05193}}].

\bibitem{Varzielas:2015iva}
I.~de~Medeiros~Varzielas and G.~Hiller, {\it {Clues for flavor from rare lepton
  and quark decays}},  {\em JHEP} {\bf 06} (2015) 072,
  [\href{http://arxiv.org/abs/1503.01084}{{\tt arXiv:1503.01084}}].

\bibitem{Kramer:2004df}
M.~Kramer, T.~Plehn, M.~Spira, and P.~Zerwas, {\it {Pair production of scalar
  leptoquarks at the CERN LHC}},  {\em Phys.Rev.} {\bf D71} (2005) 057503,
  [\href{http://arxiv.org/abs/hep-ph/0411038}{{\tt hep-ph/0411038}}].

\bibitem{CMS:zva}
{\bf CMS} Collaboration, {\it {Search for Pair-production of Second generation
  Leptoquarks in 8 TeV proton-proton collisions.}},  Tech. Rep.
  CMS-PAS-EXO-12-042, 2012.

\bibitem{Khachatryan:2014ura}
{\bf CMS} Collaboration, V.~Khachatryan et~al., {\it {Search for pair
  production of third-generation scalar leptoquarks and top squarks in
  proton-proton collisions at $\sqrt{s} = 8$ TeV}},  {\em Phys.Lett.} {\bf
  B739} (2014) 229, [\href{http://arxiv.org/abs/1408.0806}{{\tt
  arXiv:1408.0806}}].

\bibitem{Chatrchyan:2013xna}
{\bf CMS} Collaboration, S.~Chatrchyan et~al., {\it {Search for top-squark pair
  production in the single-lepton final state in pp collisions at $\sqrt{s} =
  8$ TeV}},  {\em Eur.Phys.J.} {\bf C73} (2013) 2677,
  [\href{http://arxiv.org/abs/1308.1586}{{\tt arXiv:1308.1586}}].

\bibitem{CMS-PAS-EXO-13-010}
{\bf CMS Collaboration} Collaboration, {\it {Search for Third Generation Scalar
  Leptoquarks Decaying to Top Quark - Tau Lepton Pairs in pp Collisions}},
  Tech. Rep. CMS-PAS-EXO-13-010, CERN, Geneva, 2014.

\bibitem{Aad:2011ch}
{\bf ATLAS} Collaboration, G.~Aad et~al., {\it {Search for first generation
  scalar leptoquarks in $pp$ collisions at $\sqrt{s}=7$ TeV with the ATLAS
  detector}},  {\em Phys.Lett.} {\bf B709} (2012) 158--176,
  [\href{http://arxiv.org/abs/1112.4828}{{\tt arXiv:1112.4828}}].

\bibitem{ATLAS:2012aq}
{\bf ATLAS} Collaboration, G.~Aad et~al., {\it {Search for second generation
  scalar leptoquarks in $pp$ collisions at $\sqrt{s}=7$ TeV with the ATLAS
  detector}},  {\em Eur.Phys.J.} {\bf C72} (2012) 2151,
  [\href{http://arxiv.org/abs/1203.3172}{{\tt arXiv:1203.3172}}].

\bibitem{ATLAS:2013oea}
{\bf ATLAS} Collaboration, G.~Aad et~al., {\it {Search for third generation
  scalar leptoquarks in pp collisions at $\sqrt{s}$ = 7 TeV with the ATLAS
  detector}},  {\em JHEP} {\bf 1306} (2013) 033,
  [\href{http://arxiv.org/abs/1303.0526}{{\tt arXiv:1303.0526}}].

\bibitem{Abramowicz:2012tg}
{\bf ZEUS} Collaboration, H.~Abramowicz et~al., {\it {Search for
  first-generation leptoquarks at HERA}},  {\em Phys.Rev.} {\bf D86} (2012)
  012005, [\href{http://arxiv.org/abs/1205.5179}{{\tt arXiv:1205.5179}}].

\bibitem{South:2013fta}
{\bf H1} Collaboration, D.~M. South, {\it {Search for First Generation
  Leptoquarks in ep Collisions at HERA}},  {\em PoS} {\bf ICHEP2012} (2013)
  141, [\href{http://arxiv.org/abs/1302.3378}{{\tt arXiv:1302.3378}}].

\bibitem{CMS-PAS-EXO-12-043}
{\bf CMS Collaboration} Collaboration, {\it {Search for single production of
  scalar leptoquarks in $pp$ collisions at $\sqrt{s} = 8$ TeV with the CMS
  Detector}},  Tech. Rep. CMS-PAS-EXO-12-043, CERN, Geneva, 2015.

\bibitem{Dorsner:2014axa}
I.~Dorsner, S.~Fajfer, and A.~Greljo, {\it {Cornering Scalar Leptoquarks at
  LHC}},  {\em JHEP} {\bf 1410} (2014) 154,
  [\href{http://arxiv.org/abs/1406.4831}{{\tt arXiv:1406.4831}}].

\bibitem{Mandal:2015vfa}
T.~Mandal, S.~Mitra, and S.~Seth, {\it {Single Productions of Colored Particles
  at the LHC: An Example with Scalar Leptoquarks}},  {\em JHEP} {\bf 07} (2015)
  028, [\href{http://arxiv.org/abs/1503.04689}{{\tt arXiv:1503.04689}}].

\bibitem{Khachatryan:2015vra}
{\bf CMS} Collaboration, V.~Khachatryan et~al., {\it {Searches for
  supersymmetry using the M$_{T2}$ variable in hadronic events produced in pp
  collisions at 8 TeV}},  {\em JHEP} {\bf 1505} (2015) 078,
  [\href{http://arxiv.org/abs/1502.04358}{{\tt arXiv:1502.04358}}].

\bibitem{Aad:2014wea}
{\bf ATLAS} Collaboration, G.~Aad et~al., {\it {Search for squarks and gluinos
  with the ATLAS detector in final states with jets and missing transverse
  momentum using $\sqrt{s}=8$ TeV proton--proton collision data}},  {\em JHEP}
  {\bf 1409} (2014) 176, [\href{http://arxiv.org/abs/1405.7875}{{\tt
  arXiv:1405.7875}}].

\bibitem{Aad:2015rba}
{\bf ATLAS} Collaboration, G.~Aad et~al., {\it {Search for massive, long-lived
  particles using multitrack displaced vertices or displaced lepton pairs in pp
  collisions at $\sqrt{s}$ = 8 TeV with the ATLAS detector}},
  \href{http://arxiv.org/abs/1504.05162}{{\tt arXiv:1504.05162}}.

\bibitem{Wood:1997zq}
C.~Wood, S.~Bennett, D.~Cho, B.~Masterson, J.~Roberts, et~al., {\it
  {Measurement of parity nonconservation and an anapole moment in cesium}},
  {\em Science} {\bf 275} (1997) 1759--1763.

\bibitem{Guena:2004sq}
J.~Guena, M.~Lintz, and M.~Bouchiat, {\it {Measurement of the parity violating
  $6S$-$7S$ transition amplitude in cesium achieved within $2\times 10^{-13}$
  atomic-unit accuracy by stimulated-emission detection}},  {\em Phys.Rev.}
  {\bf A71} (2005) 042108, [\href{http://arxiv.org/abs/physics/0412017}{{\tt
  physics/0412017}}].

\bibitem{Androic:2013rhu}
{\bf Qweak} Collaboration, D.~Androic et~al., {\it {First Determination of the
  Weak Charge of the Proton}},  {\em Phys.Rev.Lett.} {\bf 111} (2013), no.~14
  141803, [\href{http://arxiv.org/abs/1307.5275}{{\tt arXiv:1307.5275}}].

\bibitem{Dorsner:2011ai}
I.~Dorsner, J.~Drobnak, S.~Fajfer, J.~F. Kamenik, and N.~Kosnik, {\it {Limits
  on scalar leptoquark interactions and consequences for GUTs}},  {\em JHEP}
  {\bf 1111} (2011) 002, [\href{http://arxiv.org/abs/1107.5393}{{\tt
  arXiv:1107.5393}}].

\bibitem{Minkowski:1977sc}
P.~Minkowski, {\it {$\mu \to e\gamma$ at a Rate of One Out of $10^{9}$ Muon
  Decays?}},  {\em Phys.Lett.} {\bf B67} (1977) 421--428.

\bibitem{Yanagida:1979as}
T.~Yanagida, {\it {HORIZONTAL SYMMETRY AND MASSES OF NEUTRINOS}},  {\em
  Conf.Proc.} {\bf C7902131} (1979) 95--99.

\bibitem{GellMann:1980vs}
M.~Gell-Mann, P.~Ramond, and R.~Slansky, {\it {Complex Spinors and Unified
  Theories}},  {\em Conf.Proc.} {\bf C790927} (1979) 315--321,
  [\href{http://arxiv.org/abs/1306.4669}{{\tt arXiv:1306.4669}}].

\bibitem{Glashow:1979nm}
S.~Glashow, {\it {The Future of Elementary Particle Physics}},  {\em NATO
  Sci.Ser.B} {\bf 59} (1980) 687.

\bibitem{Mohapatra:1979ia}
R.~N. Mohapatra and G.~Senjanovic, {\it {Neutrino Mass and Spontaneous Parity
  Violation}},  {\em Phys.Rev.Lett.} {\bf 44} (1980) 912.

\bibitem{Asaka:2005an}
T.~Asaka, S.~Blanchet, and M.~Shaposhnikov, {\it {The nuMSM, dark matter and
  neutrino masses}},  {\em Phys.Lett.} {\bf B631} (2005) 151--156,
  [\href{http://arxiv.org/abs/hep-ph/0503065}{{\tt hep-ph/0503065}}].

\bibitem{Asaka:2005pn}
T.~Asaka and M.~Shaposhnikov, {\it {The nuMSM, dark matter and baryon asymmetry
  of the universe}},  {\em Phys.Lett.} {\bf B620} (2005) 17--26,
  [\href{http://arxiv.org/abs/hep-ph/0505013}{{\tt hep-ph/0505013}}].

\bibitem{Boyarsky:2009ix}
A.~Boyarsky, O.~Ruchayskiy, and M.~Shaposhnikov, {\it {The Role of sterile
  neutrinos in cosmology and astrophysics}},  {\em Ann.Rev.Nucl.Part.Sci.} {\bf
  59} (2009) 191--214, [\href{http://arxiv.org/abs/0901.0011}{{\tt
  arXiv:0901.0011}}].

\bibitem{Kuzmin:1985mm}
V.~Kuzmin, V.~Rubakov, and M.~Shaposhnikov, {\it {On the Anomalous Electroweak
  Baryon Number Nonconservation in the Early Universe}},  {\em Phys.Lett.} {\bf
  B155} (1985) 36.

\bibitem{Olive:1981ak}
K.~A. Olive and M.~S. Turner, {\it {Cosmological Bounds on the Masses of
  Stable, Right-handed Neutrinos}},  {\em Phys.Rev.} {\bf D25} (1982) 213.

\bibitem{Peebles:1982ib}
P.~Peebles, {\it {PRIMEVAL ADIABATIC PERTURBATIONS: EFFECT OF MASSIVE
  NEUTRINOS}},  {\em Astrophys.J.} {\bf 258} (1982) 415--424.

\bibitem{Pal:1981rm}
P.~B. Pal and L.~Wolfenstein, {\it {Radiative Decays of Massive Neutrinos}},
  {\em Phys.Rev.} {\bf D25} (1982) 766.

\bibitem{Barger:1995ty}
V.~D. Barger, R.~Phillips, and S.~Sarkar, {\it {Remarks on the KARMEN
  anomaly}},  {\em Phys.Lett.} {\bf B352} (1995) 365--371,
  [\href{http://arxiv.org/abs/hep-ph/9503295}{{\tt hep-ph/9503295}}].

\bibitem{Dolgov:2000ew}
A.~Dolgov and S.~Hansen, {\it {Massive sterile neutrinos as warm dark matter}},
   {\em Astropart.Phys.} {\bf 16} (2002) 339--344,
  [\href{http://arxiv.org/abs/hep-ph/0009083}{{\tt hep-ph/0009083}}].

\bibitem{Abazajian:2001vt}
K.~Abazajian, G.~M. Fuller, and W.~H. Tucker, {\it {Direct detection of warm
  dark matter in the X-ray}},  {\em Astrophys.J.} {\bf 562} (2001) 593--604,
  [\href{http://arxiv.org/abs/astro-ph/0106002}{{\tt astro-ph/0106002}}].

\bibitem{Shi:1998km}
X.-D. Shi and G.~M. Fuller, {\it {A New dark matter candidate: Nonthermal
  sterile neutrinos}},  {\em Phys.Rev.Lett.} {\bf 82} (1999) 2832--2835,
  [\href{http://arxiv.org/abs/astro-ph/9810076}{{\tt astro-ph/9810076}}].

\bibitem{Akhmedov:1998qx}
E.~K. Akhmedov, V.~Rubakov, and A.~Y. Smirnov, {\it {Baryogenesis via neutrino
  oscillations}},  {\em Phys.Rev.Lett.} {\bf 81} (1998) 1359--1362,
  [\href{http://arxiv.org/abs/hep-ph/9803255}{{\tt hep-ph/9803255}}].

\bibitem{Shaposhnikov:2008pf}
M.~Shaposhnikov, {\it {The nuMSM, leptonic asymmetries, and properties of
  singlet fermions}},  {\em JHEP} {\bf 0808} (2008) 008,
  [\href{http://arxiv.org/abs/0804.4542}{{\tt arXiv:0804.4542}}].

\bibitem{Canetti:2010aw}
L.~Canetti and M.~Shaposhnikov, {\it {Baryon Asymmetry of the Universe in the
  NuMSM}},  {\em JCAP} {\bf 1009} (2010) 001,
  [\href{http://arxiv.org/abs/1006.0133}{{\tt arXiv:1006.0133}}].

\bibitem{Asaka:2010kk}
T.~Asaka and H.~Ishida, {\it {Flavour Mixing of Neutrinos and Baryon Asymmetry
  of the Universe}},  {\em Phys.Lett.} {\bf B692} (2010) 105--113,
  [\href{http://arxiv.org/abs/1004.5491}{{\tt arXiv:1004.5491}}].

\bibitem{Asaka:2011wq}
T.~Asaka, S.~Eijima, and H.~Ishida, {\it {Kinetic Equations for Baryogenesis
  via Sterile Neutrino Oscillation}},  {\em JCAP} {\bf 1202} (2012) 021,
  [\href{http://arxiv.org/abs/1112.5565}{{\tt arXiv:1112.5565}}].

\bibitem{Canetti:2012vf}
L.~Canetti, M.~Drewes, and M.~Shaposhnikov, {\it {Sterile Neutrinos as the
  Origin of Dark and Baryonic Matter}},  {\em Phys.Rev.Lett.} {\bf 110} (2013)
  061801, [\href{http://arxiv.org/abs/1204.3902}{{\tt arXiv:1204.3902}}].

\bibitem{Canetti:2012zc}
L.~Canetti, M.~Drewes, and M.~Shaposhnikov, {\it {Matter and Antimatter in the
  Universe}},  {\em New J.Phys.} {\bf 14} (2012) 095012,
  [\href{http://arxiv.org/abs/1204.4186}{{\tt arXiv:1204.4186}}].

\bibitem{Canetti:2012kh}
L.~Canetti, M.~Drewes, T.~Frossard, and M.~Shaposhnikov, {\it {Dark Matter,
  Baryogenesis and Neutrino Oscillations from Right Handed Neutrinos}},  {\em
  Phys.Rev.} {\bf D87} (2013) 093006,
  [\href{http://arxiv.org/abs/1208.4607}{{\tt arXiv:1208.4607}}].

\bibitem{Pilaftsis:2003gt}
A.~Pilaftsis and T.~E. Underwood, {\it {Resonant leptogenesis}},  {\em
  Nucl.Phys.} {\bf B692} (2004) 303--345,
  [\href{http://arxiv.org/abs/hep-ph/0309342}{{\tt hep-ph/0309342}}].

\bibitem{Wolfenstein:1977ue}
L.~Wolfenstein, {\it {Neutrino Oscillations in Matter}},  {\em Phys.Rev.} {\bf
  D17} (1978) 2369--2374.

\bibitem{Arcadi:2014dca}
G.~Arcadi, L.~Covi, and F.~Dradi, {\it {3.55 keV line in Minimal Decaying Dark
  Matter scenarios}},  \href{http://arxiv.org/abs/1412.6351}{{\tt
  arXiv:1412.6351}}.

\bibitem{Bulbul:2014sua}
E.~Bulbul, M.~Markevitch, A.~Foster, R.~K. Smith, M.~Loewenstein, et~al., {\it
  {Detection of An Unidentified Emission Line in the Stacked X-ray spectrum of
  Galaxy Clusters}},  {\em Astrophys.J.} {\bf 789} (2014) 13,
  [\href{http://arxiv.org/abs/1402.2301}{{\tt arXiv:1402.2301}}].

\bibitem{Boyarsky:2014jta}
A.~Boyarsky, O.~Ruchayskiy, D.~Iakubovskyi, and J.~Franse, {\it {Unidentified
  Line in X-Ray Spectra of the Andromeda Galaxy and Perseus Galaxy Cluster}},
  {\em Phys.Rev.Lett.} {\bf 113} (2014) 251301,
  [\href{http://arxiv.org/abs/1402.4119}{{\tt arXiv:1402.4119}}].

\bibitem{Mambrini:2013iaa}
Y.~Mambrini, K.~A. Olive, J.~Quevillon, and B.~Zaldivar, {\it {Gauge Coupling
  Unification and Nonequilibrium Thermal Dark Matter}},  {\em Phys.Rev.Lett.}
  {\bf 110} (2013) 241306, [\href{http://arxiv.org/abs/1302.4438}{{\tt
  arXiv:1302.4438}}].

\bibitem{Georgi:1974yf}
H.~Georgi, H.~R. Quinn, and S.~Weinberg, {\it {Hierarchy of Interactions in
  Unified Gauge Theories}},  {\em Phys.Rev.Lett.} {\bf 33} (1974) 451--454.

\bibitem{Machacek:1983tz}
M.~E. Machacek and M.~T. Vaughn, {\it {Two Loop Renormalization Group Equations
  in a General Quantum Field Theory. 1. Wave Function Renormalization}},  {\em
  Nucl.Phys.} {\bf B222} (1983) 83.

\bibitem{Murayama:1991ah}
H.~Murayama and T.~Yanagida, {\it {A viable SU(5) GUT with light leptoquark
  bosons}},  {\em Mod.Phys.Lett.} {\bf A7} (1992) 147--152.

\bibitem{moriondsuperk}
S.~Mine, ``{Recent results from SuperK}.''
  \url{https://indico.in2p3.fr/event/10819/session/0/contribution/81/material/slides/0.pdf}.
\newblock talk presented at 50th Rencontres de Moriond EW 2015, March 14--21,
  2015.

\bibitem{Ellis:2015jwa}
S.~A.~R. Ellis and J.~D. Wells, {\it {Visualizing gauge unification with
  high-scale thresholds}},  {\em Phys.Rev.} {\bf D91} (2015) 075016,
  [\href{http://arxiv.org/abs/1502.01362}{{\tt arXiv:1502.01362}}].

\bibitem{Sakai:1981pk}
N.~Sakai and T.~Yanagida, {\it {Proton Decay in a Class of Supersymmetric Grand
  Unified Models}},  {\em Nucl.Phys.} {\bf B197} (1982) 533.

\bibitem{Weinberg:1981wj}
S.~Weinberg, {\it {Supersymmetry at Ordinary Energies. 1. Masses and
  Conservation Laws}},  {\em Phys.Rev.} {\bf D26} (1982) 287.

\bibitem{Dorsner:2005fq}
I.~Dorsner and P.~Fileviez~Perez, {\it {Unification without supersymmetry:
  Neutrino mass, proton decay and light leptoquarks}},  {\em Nucl.Phys.} {\bf
  B723} (2005) 53--76, [\href{http://arxiv.org/abs/hep-ph/0504276}{{\tt
  hep-ph/0504276}}].

\bibitem{Affleck:1984fy}
I.~Affleck and M.~Dine, {\it {A New Mechanism for Baryogenesis}},  {\em
  Nucl.Phys.} {\bf B249} (1985) 361.

\bibitem{Dutta:2015dka}
B.~Dutta, Y.~Gao, T.~Li, C.~Rott, and L.~E. Strigari, {\it {Leptoquark
  implication from the CMS and IceCube experiments}},  {\em Phys.Rev.} {\bf
  D91} (2015) 125015, [\href{http://arxiv.org/abs/1505.00028}{{\tt
  arXiv:1505.00028}}].

\end{thebibliography}\endgroup


\end{document}